\documentclass[%
aip,
superscriptaddress,
amsmath,amssymb,
reprint,%
]{revtex4-2}

\usepackage{graphicx}
\usepackage{dcolumn}
\usepackage{bm}

\usepackage[utf8]{inputenc}
\usepackage[T1]{fontenc}
\usepackage{mathptmx}
\usepackage{etoolbox}
\usepackage{xcolor}
\usepackage{verbatim}
\usepackage{textcomp}
\usepackage[american]{babel}

\makeatletter
\def\@email#1#2{%
	\endgroup
	\patchcmd{\titleblock@produce}
	{\frontmatter@RRAPformat}
	{\frontmatter@RRAPformat{\produce@RRAP{*#1\href{mailto:#2}{#2}}}\frontmatter@RRAPformat}
	{}{}
}%
\makeatother

\begin{document}

	\title[Numerical analysis of a 3WM JTWPA with
		engineered dispersion loadings]{Numerical analysis of a three-wave-mixing Josephson traveling-wave parametric amplifier with
		engineered dispersion loadings}
				
	\author{Victor Gaydamachenko}
    \email{victor.gaydamachenko@ptb.de, christoph.kissling@ptb.de}
	
	\affiliation{Physikalisch-Technische Bundesanstalt, Bundesallee
		100, 38116 Braunschweig, Germany}	
	
	\author{Christoph Kissling}
	\affiliation{Physikalisch-Technische Bundesanstalt, Bundesallee
		100, 38116 Braunschweig, Germany}    
	
	\author{Ralf Dolata}
	\affiliation{Physikalisch-Technische Bundesanstalt, Bundesallee
		100, 38116 Braunschweig, Germany}
	\author{Alexander B. Zorin}
	\affiliation{Physikalisch-Technische Bundesanstalt, Bundesallee
		100, 38116 Braunschweig, Germany}
	
	\date{\today}

	\begin{abstract}
        
    		The recently proposed Josephson traveling-wave  parametric amplifier (JTWPA)
    		based on a ladder transmission line consisting of
    		radio-frequency SQUIDs and exploiting three-wave mixing (3WM), has great potential in achieving both a gain of 20 dB and a flat bandwidth of at least 4 GHz.
    		To realize this concept in practical amplifiers we model
    		the advanced JTWPA circuit with periodic modulation of the
    		circuit parameters (engineered dispersion loadings),
    		which allow the basic mixing process, i.e., $\omega_s=\omega_p-\omega_i$, where $\omega_s$, $\omega_p$, and
    		$\omega_i$ are the signal, the pump, and the idler frequencies, respectively, and efficiently suppress propagation of
    		unwanted higher tones including $\omega_{2p}=2\omega_p$, $\omega_{p+s}=\omega_p +\omega_s$, $\omega_{p+i} = \omega_p + \omega_i$, etc. The engineered dispersion loadings 
            allow achieving sufficiently wide $3$~dB-bandwidth from $3$~GHz to $9$~GHz combined with a reasonably small ripple ($\pm2$~dB) in the gain-versus-frequency dependence.
		
	\end{abstract}
	\maketitle
	
	\section{Introduction}
	
	Due to attainable quantum-limited performance, Josephson parametric amplifiers (JPAs)
	make a multitude of microwave quantum experiments possible, e.g., amplification and squeezing of zero-point fluctuations
	\cite{Movshovich1990,Castellanos-Beltran2008}, dramatic improvement of sensitivity
	of quantum sensors \cite{Hatridge2011}, observation of quantum jumps
	\cite{Vijay2011}, generation of entangled microwave radiation \cite{Flurin2012},
	single-shot \cite{Lin2013} and continuous nondemolition \cite{Vool2016}
	measurements of superconducting qubits, etc. Nowadays they
	are considered as indispensable tools for quantum technologies \cite{Devoret2013}.
	Recently, Josephson \emph{traveling-wave} parametric amplifiers (JTWPAs) \cite{Yaakobi2013, OBrien2014, White2015, Macklin2015, Bell-Samolov2015, Zorin2016,Zorin2017, WenyuanZhang2017, Miano2018, Zorin2019, Planat2020, Ranadive2022} are in particular focus of
	research in the fields of quantum communication and quantum computing.
	Thanks to their transmission-line architecture, leaving off a customary cavity, 
	JTWPAs achieve a substantially wider bandwidth and larger dynamic range \cite{OBrien2014}.
	
	The core of a JTWPA is a nonlinear microwave transmission line, made up of a LC-ladder
	circuit with cells, which include either single
	Josephson junctions (see, e.g., Refs. \cite{Yaakobi2013,OBrien2014,Macklin2015,White2015}) or SQUIDs
	(see, e.g., Refs. \cite{Bell-Samolov2015,Zorin2016,Zorin2017,WenyuanZhang2017,Miano2018,Zorin2019,Planat2020, Ranadive2022,Zorin2021}).
	Due to their nonlinear current-phase relation (or, equivalently, nonlinear
	Josephson inductance $L_J$) these elements enable frequency mixing. 
	The conventional scheme of
	JTWPAs \cite{Yaakobi2013,OBrien2014,Macklin2015,White2015,Bell-Samolov2015,Planat2020, Ranadive2022}
	is based on four-wave mixing (4WM),
	where frequencies of the signal, $\omega_s$, the pump, $\omega_p$, and the idler, $\omega_i$,
	obey the relation $\omega_s + \omega_i = 2\omega_p$.
	Such mixing is enabled by the Kerr-like nonlinearity of the inverse Josephson
	inductance, $L_J^{-1}(\phi) \approx (1-\gamma'\phi^2) L_{J0}^{-1}$ with coefficient $\gamma'=1/6$, which
	stems from the Taylor-series expansion of the sine-shape current-phase relation
	of the Josephson current, $I = I_c \sin \phi$ \cite{Josephson1962}. Here $I_c$ is the critical current,
	$L_{J0}^{-1} = \left[\partial I(\phi)/\partial (\varphi_0\phi)\right]_{\phi=0} = I_c/\varphi_0$ is
	the inverse Josephson inductance for vanishingly small signals, and $\varphi_0 = \hbar/2e$
	is the reduced magnetic flux quantum.
	
	The major challenge in designing JTWPAs with 4WM is fulfilling the matching relation
	for the corresponding wave numbers, $k_s + k_i = 2 k_p$, in the range of operating
	frequencies and powers. The problem originates from the properties of the Kerr effect
	which, on the one hand, ensures 4WM, but, on the other hand, causes unwanted
	self-phase modulation (SPM) and cross-phase modulation (XPM)
	of the waves due to the intensity-dependent refractive index of the line. This effect
	leads to imperfect phase matching, $\Delta k = 2k_p-k_s-k_i \neq 0$, which in addition
	is very sensitive to pump power \cite{Agrawal}.
	This problem of amplifiers with 4WM can be fixed by careful dispersion engineering, either by applying the resonant phase matching technique \cite{OBrien2014,Macklin2015,White2015}, where resonators are inserted into the transmission line at regular intervals, or by using periodic loadings in the transmission line, opening stop-bands in the dispersion relation \cite{Planat2020}. In this way the phase velocity of the pump wave is adjusted yielding $\Delta k \rightarrow 0$
	However, both techniques lead to a wide stop-band in the center of the gain profile. A recently demonstrated approach circumvents the need of dispersion engineering by reversing the sign of the Kerr nonlinearity, such that the nonlinear Kerr-induced phase shift compensates the linear dispersion phase mismatch \cite{Bell-Samolov2015, Ranadive2022}. Although this elegant approach overcomes the problem of a stop-band in the gain profile, its gain profile is rather camel-back shaped than being flat.
		
	Recently, the concept of a JTWPA with three-wave-mixing (3WM) was proposed \cite{Zorin2016}
	and tested \cite{Zorin2017, Miano2018}.
	In amplifiers of this type, the frequencies
	obey the basic parametric relation, $\omega_s+\omega_i = \omega_p$,
	and their mixing is possible due to the non-centrosymmetric nonlinearity 
	of type \cite{Tien1958,Cullen1960}
	\begin{equation}
		L_S^{-1}(\phi) \approx (1 - \beta' \phi)L_{S0}^{-1},
		\label{L-3ph}
	\end{equation}
	where $\beta'$ is the coefficient of the non-centrosymmetric nonlinearity, and $L_S$ and $L_{S0}$ denote the inductance for finite and vanishingly small signals, respectively.
	Such a nonlinearity can be engineered, for example, with the help
	of a flux-biased non-hysteretic rf-SQUID \cite{Zorin2016}.
	Alternatively, this nonlinearity can be engineered by using a modified (multi-junction) rf-SQUID having a Josephson kinetic inductance
	instead of linear inductance in one branch, i.e., the so-called superconducting nonlinear
	asymmetric inductive element (SNAIL) \cite{Sivak2019,Zorin2017,Frattini2017,Frattini2018}.
	Fine tuning of the bias flux allows full suppression of the Kerr nonlinearity in both rf SQUIDs and SNAILs. 
	
	Compared to the Kerr-based JTWPA with 4WM, the JTWPA with 3WM normally contains only a small or even vanishing Kerr-like nonlinearity, so it is almost free of SPM and XPM effects \cite{Agrawal} and the corresponding wave numbers
	$k_s$, $k_i$ and $k_p$ do not depend on the pump power.
	Although this remarkable property intrinsically ensures phase-matching for the basic 3WM process, 	\begin{equation}
		\Delta k = k_p-k_s-k_i \approx 0
		\label{Eq-PhaseMatching3WM}
	\end{equation} (naturally, in case of sufficiently small chromatic dispersion) \cite{Zorin2016}, the phase-matching conditions for unwanted mixing processes with higher frequencies are also satisfied. These processes include, e.g., the generation of the second harmonic of the pump, $\omega_{2p}=2\omega_p$, and the sum-frequency generations, $\omega_{p+s}=\omega_p+\omega_s$ and $\omega_{p+i}=\omega_p+\omega_i$. All these unwanted mixing processes cause leakage of the pump power \cite{White2015,Zorin2016} and dramatically limit the achievable signal gain \cite{Dixon2020}. In fact, the amplitudes of these tones can be of the same order as pump and signal, respectively \cite{Dixon2020}. This problem can be solved by engineering a relatively strong dispersion in the transmission line to break the phase-matching condition for all unwanted processes \cite{Zorin2021, Malnou2021, Perelshtein21}. This can be done, for example, by lowering the SQUID plasma frequency \cite{Zorin2019, Zorin2021}.  At the same time, the phase-matching of the relevant 3WM process can be retained by techniques like resonant-phase-matching \cite{OBrien2014} or quasi-phase-matching \cite{Zorin2021}. In this paper, however, we show how periodic dispersion loadings in the transmission line can be engineered such that the problem of unwanted mixing processes is solved and simultaneously the bandwidth and flatness of the resulting gain profile is maximized without change of the circuit architecture.

	\section{Circuit description}

	\begin{figure}[b]
		\begin{center}
			\includegraphics[width=8.6cm]{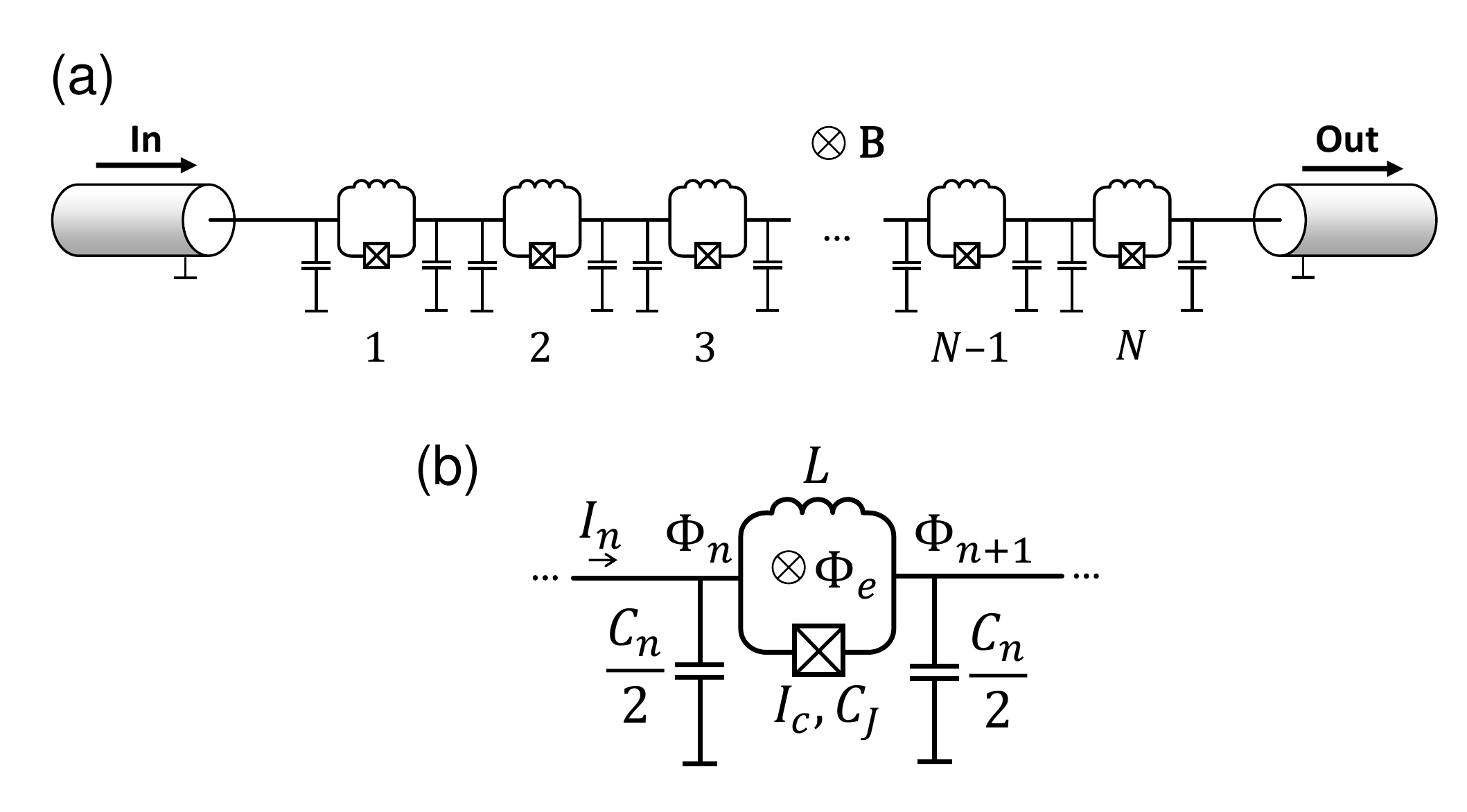}
			\caption{(a)~Electric diagram of the JTWPA consisting of a serial array of $N \gg 1$
				identical rf-SQUIDs with not necessarily identical ground capacitances $C_n$.
				Homogeneous magnetic field $\textbf{B}$ is perpendicular to the circuit plane and produces, as shown in panel (b), similar constant flux $\Phi_e$ applied to each rf-SQUID having  inductance $L$ and a Josephson junction with critical current $I_c$ and self-capacitance $C_J$.
				The basic circuit variables are $\Phi_n$, the flux value on the $n$-th node of the line, and $I_n$, the alternating current generated by a microwave and injected in the $n$-th rf-SQUID. 
				} 
				\label{fig-EqvSchm}
		\end{center}
	\end{figure}

	The electric diagram of our superconducting microwave circuit with nominally vanishing losses is shown in Fig.~\ref{fig-EqvSchm}a.
	The circuit consists of an array of $N$ elementary cells (see Fig.~\ref{fig-EqvSchm}b). Each elementary $\pi$-type unit cell consists of an rf-SQUID, deployed as a nonlinear inductance enabling 3WM \cite{Zorin2016}, and two capacitances to ground $0.5 C_n$ (with integer $n=1...N$), which are periodically varied to create the dispersion loadings. 
    All $N$ rf-SQUIDs, consisting of an inductance $L$ in parallel to a Josephson junction having critical current $I_c$ and junction capacitance $C_J$, are nominally identical and nonhysteretic, i.e., the dimensionless screening parameter $\beta_L \equiv LI_c/\varphi_0 < 1$  \cite{ClarkeBraginski2004}. 
	
	For a vanishingly small ac current $I$ injected in the rf-SQUID the \textit{linear}
	inductance equals
	\begin{equation}
		L_S(I \rightarrow 0) \equiv L_{S0} = L /\left(1 + \beta_L\cos \phi_{\textrm{dc}}\right),
		\label{L_S}
	\end{equation}
	where the phase $\phi_{\textrm{dc}}$ is set by an external magnetic flux $\Phi_e$ \cite{Zorin2016}. 
	Thus, the inductance of the rf-SQUIDs can be effectively controlled \textit{in situ} by a magnetic field $\textbf{B}$, 
	which is applied equally to all rf-SQUIDs, or alternatively, by injecting a dc bias current $I_{\textrm{dc}}$ into the transmission line, and thus inducing $\Phi_e=LI_{\textrm{dc}}$.

	Considering now a small but not vanishing ac phase perturbation $\phi=(\Phi_n-\Phi_{n+1})/\varphi_0$ on the $n$-th rf-SQUID, induced by small injected ac current $I\ll \Phi_0/L_{S0}$, we find the inverse \textit{nonlinear} SQUID inductance
	\begin{equation}
		L_S^{-1} = \frac{\partial I(\phi)}{\partial (\phi\varphi_0)} \approx (1 - 2\beta \phi - 3\gamma \phi^2) L_{S0}^{-1}, 
		\label{L-vs-phi}
	\end{equation}
	with $\beta$ and $\gamma$ denoting the coefficients of the non-centrosymmetric and the Kerr-nonlinearity, respectively.  
	The coefficients  $\beta$ and $\gamma$, 
	\begin{eqnarray}
		\beta  = \frac{\beta_L}{2}\frac{\sin \phi_{\textrm{dc}}}{1+\beta_L \cos \phi_{\textrm{dc}}},
		\label{bet-gam-finitebetaL1} \\
		\gamma  = \frac{\beta_L}{6}\frac{\cos \phi_{\textrm{dc}}}{1+\beta_L \cos \phi_{\textrm{dc}}},
		\label{bet-gam-finitebetaL2}
	\end{eqnarray}
	are odd and even $2\pi$-periodic functions of the phase $\phi_{\textrm{dc}}$, respectively. 
	Thus, the nonlinearity of the SQUID inductance is effectively controllable by $\Phi_e$, too.
	Since 3WM is enabled by the non-centrosymmetric nonlinearity, we aim at a relatively large value of $\beta$. 
	To reduce unwanted Kerr effects causing SPM and XPM,  $\gamma$ should be small. 
	In this paper, we bias the SQUIDs by $I_\textrm{dc}$ such, that  
	$\phi_\textrm{dc}\approx \pi/2+\beta_L$, for which $\beta$ is close to its maximum value, and $\gamma$ is sufficiently small, so that SPM and XPM do not cause substantial phase mismatch.

	The nominal value of the inductance $L$ and the average value of the ground capacitances 
	\begin{equation}
	    \overline{C} =\frac{1}{m}\sum_{n = 1}^{m} C_n,
	    \label{average_capacitance}
	\end{equation} 
	are chosen such that the characteristic impedance of the equivalent uniform transmission line for sufficiently low frequencies,
	\begin{equation}
		Z \approx \overline{Z}=\sqrt{\frac{L_{S0}}{\overline{C}}} = \sqrt{\frac{L}{\overline{C}(1 + \beta_L\cos\phi_{\textrm{dc})}}},
		\label{lineimpedance}
	\end{equation}
	is equal to $Z_0=50\,\Omega$ at the envisaged operation point $\phi_\textrm{dc}$ to achieve impedance-matching to the amplifier's experimental environment. Note that this impedance can be in-situ flux-tuned.
	
    The self-capacitance $C_J$ of the Josephson junction yields the plasma frequency of the rf SQUID $\omega_J=(L_{S0}C_J)^{-1/2}$, and the ground capacitances define the characteristic frequency $\omega_0=(L_{S0}\overline{C})^{-1/2}$, both constituting the cutoff frequency\cite{Dixon2020}
    \begin{equation}
		\omega_c=2(L_{S0}(\overline{C}+4C_J))^{-1/2}=(\omega_0^{-2}/4+\omega_J^{-2})^{-1/2}
		\label{eq-cutoff}
	\end{equation} 
    No propagation of electromagnetic waves along the transmission line is possible for $\omega>\omega_c$. 
    For lower frequencies, the transmission line becomes dispersive, and the dispersion relation $k(\omega)$ is approximately
    \begin{equation}
		k \approx  \frac{\omega}{\omega_0} \left(1+ \frac{\omega^2}{2\omega_J^2}
		+ \frac{\omega^2}{24\omega_0^2} \right)
		\label{k-vs-low-w}
	\end{equation}
    for frequencies $\omega \ll \omega_0, \omega_J$.
    Here, the wave number $k$ is normalized to the reverse physical size
	of the elementary cell $d^{-1}$ ($d = \ell/N$, where $\ell$ is the
	total length of the line).
	Since typical signal, idler, and pump frequencies (e.g., $\omega_{s,i}/2\pi \approx 4...8\;\textrm{GHz}$ and $\omega_p/2\pi\approx  12\;\textrm{GHz}$) are much smaller than both the frequencies $\omega_{0,J}/2\pi\approx  80...100\;\textrm{GHz}$, the dispersion relation is roughly linear, $k\approx \omega/\omega_0$. 
    Therefore, for the 3WM process, with $\omega_p=\omega_s+\omega_i$, the phase matching condition $k_p\approx k_s+k_i$ is approximately fulfilled. However, due to the almost linear  dispersion relation, the phase matching conditions for \textit{unwanted} 3WM processes, e.g.,
    \begin{align}
		\omega_{p+s} &= \omega_p + \omega_s\,,& k_{p+s} &\approx k_p + k_s \,,  \label{eq-w1} \\
		\omega_{p+i} &= \omega_p + \omega_i\,,& k_{p+i} &\approx k_p + k_i  \,,  \label{eq-w2}\\
		\omega_{2p} &= 2\omega_p\,,& k_{2p} &\approx 2 k_p \,,  \label{eq-w2p}
	\end{align}
    are roughly fulfilled, too. 
    These processes are unwanted since they lead
    to leakage of power and  undulations of the signal gain \cite{Zorin2019,Zhao2021} (c.f. Appendix~\ref{appendix_nogap}).
    It was shown recently via circuit simulations \cite{Dixon2020} that indeed a large number of unwanted mixing processes take place in a 3WM-JTWPA with a homogeneous transmission line and relatively low chromatic dispersion ($f_J = \omega_J / 2\pi = 86$~GHz, $f_0 = \omega_0 / 2\pi = 67$~GHz), and that the gain in such an amplifier was thereby limited to rather smaller values on the order of 10 dB.

	\begin{figure*}[t]
		\begin{center}
			\includegraphics[width=7in]{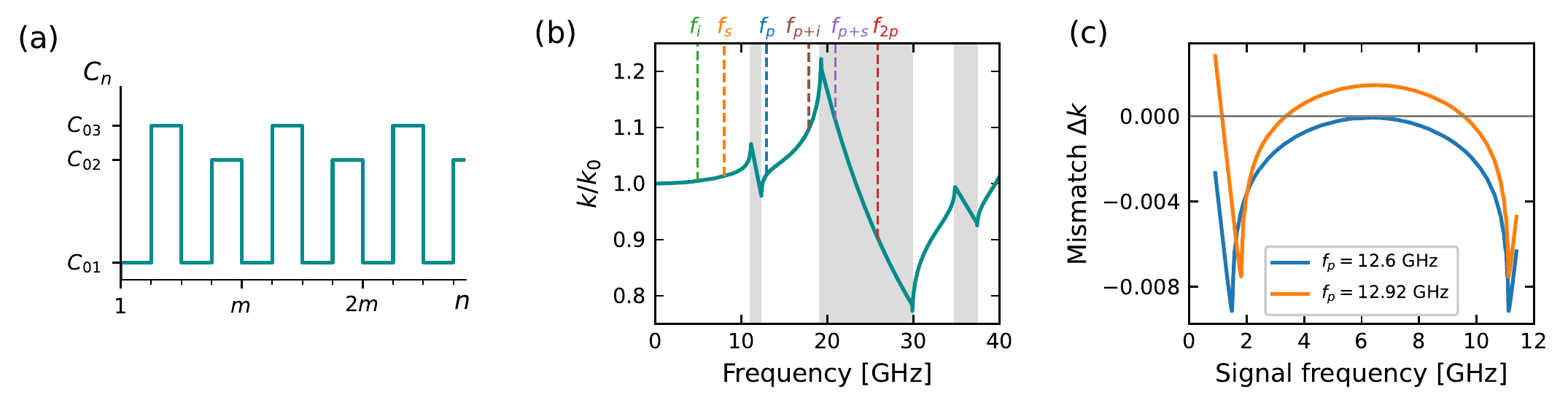}
			\caption{Dispersion engineering via periodic loadings. (a)~Scheme of periodic variation of the ground capacitance $C_n$ over the length of the JTWPA transmission line. 
			The values of ground capacitances $C_n$, $n = 1, 2,..., N$, are altered between three constant values, with the global period  $m$. 
			It is assumed that $0.25 m$ is an integer and, for example, the first segment with $C_n = C_{01}$ extends from $n =1$ up to $n = 0.25 m$ inclusively. (b)~Resulting dispersion relation, normalized on  $k_0=\omega/\omega_0$. Grey-shaded areas denote stop-bands (gaps), opened due to the periodic loadings shown in panel~(a). 
			The pump frequency is fixed at the upper edge of the first gap (corresponding to loading period $m$) so that $k_p$ is lowered and phase-matching of the pump, the signal, and the idler is achieved.  
			All unwanted higher-frequency tones, $f_{p+s},f_{p+i}, f_{2p}$, etc., are strongly phase-mismatched since they fall either into the highly dispersive region above $f_p$ or into the wide second gap (corresponding to loading period $m/2$). 
			The position of the six most relevant 3WM tones are marked by vertical dashed lines for the case of $f_p=12.92~\textrm{GHz}$ and $f_s=8~\textrm{GHz}$.  (c)~Phase-mismatch $\Delta k=k_p-k_s-k_i$ versus $f_s$ for two different pump frequencies.   
			Plots in (b) and (c) are calculated using TMM for the circuit parameters $C_{01} = 8.8$~fF, $C_{02} = 62.3$~fF, $C_{03} = 80$~fF, $m = 20$, $L_{S0}=109$~pH, $C_{J} = 20$~fF, and $N=1500$. 
			} \label{fig-profile}
		\end{center}
	\end{figure*}
	
	\section{Dispersion engineering}

    It is the major task in the design of a 3WM-JTWPA to destroy the phase-matching in all unwanted processes Eq.~(\ref{eq-w1})-(\ref{eq-w2p}), while preserving it solely for the basic 3WM process Eq.~(\ref{Eq-PhaseMatching3WM}). 
	Our approach to tackle this issue is based on periodic dispersion loadings, which has the advantage that no modification of the circuit architecture is required (like, for example, inserting resonator-based phase-shifters into the transmission-line  \cite{White2015,Macklin2015}), but only a variation of the circuit parameters is needed.
     
	The method of periodic dispersion loading has earlier been applied for 4WM traveling-wave amplifiers, based on the nonlinearity of the kinetic inductance (KI-TWPAs) and designed as a superconducting coplanar waveguide transmission line \cite{Eom2012}. 
	The engineered loadings were realized as waveguide regions of certain length with increased width. 
	Placed at specific intervals, these loadings formed a frequency stop band and thus prevented propagation of unwanted waves (e.g., the third harmonic of the pump generated due to the Kerr nonlinearity). 
	KI-TWPAs with additional dc biasing enabling both 4WM and 3WM \cite{Vissers2016, Malnou2021} were realized using similar concepts. 
	In their analysis of periodic loadings in KI-TWPA,
	Erickson and Pappas \cite{Erickson2017} suggested that this method
	may also be applicable to the lumped-element (Josephson-junction based)
	traveling-wave parametric amplifiers. This concept was demonstrated recently by Planat et al. \cite{Planat2020} for a 4WM-JTWPA. In this paper we extend the concept of periodic loadings to the case of a 3WM-JTWPA based on an rf-SQUID array (c.f. Refs.\cite{Roudsari_arxiv2022, Nillson_arxiv2022}).

    In our approach, the periodic loadings are realized by variation of ground capacitances $C_n$. 
    The periodic variation of the ground capacitances leads to a corresponding modulation of the local value of phase velocity, $v_n = d/\sqrt{L_{S0}C_n}$, with a shape similar to that in the Kronig-Penney model \cite{KronigPenney1931}. The wave frequency $\omega(k)$, as a function of (Bloch) wave number $k$, forms in this case a band structure with opened gaps (frequency bands of forbidden wave propagation) at $k_j = jk_m/2$, where $k_m = 2\pi/m$, and centered around frequencies $\omega_j = (j\pi/m)\omega_0$, where integer $m$ is the period of the capacitance variation and integer variable $j = $ 1, 2, etc.     The size of the $j$-th gap is proportional to the corresponding $j$-th Fourier coefficient in the expansion of the Kronig-Penney potential \cite{Kittel2004}.  
    
    Our periodic loadings are designed such that the first gap ($j=1$) of the resulting band structure $\omega(k)$ is rather narrow and the second gap ($j=2$) is wide. This is achieved using a
	variation of the ground capacitances $C_n$ with three discrete capacitances as shown in Fig.~\ref{fig-profile}a.
	This modified Kronig-Penney waveform \cite{KronigPenney1931} has a basic period of $m$ or, equivalently,
	the dimensionless space frequency of $k_m/2\pi=1/m$. However, the double-frequency
	component $2/m$ of this waveform  (corresponding to the period of $m/2$) has
	substantially larger modulation depth than that of the basic frequency $1/m$.
	This relation is due to inserting two slightly different capacitances for each half-period.
	Therefore, the width of the second gap in the spectrum
	$\omega(k)$ is expected to be notably larger than that of the first gap.  The shape of  $\omega(k)$ can be calculated analytically using the transfer-matrix-method (TMM) \cite{Mackay2020}, as described in Appendix~\ref{appendix-TMM}, and is presented in Fig.~\ref{fig-profile}b.

	The purpose of the narrow first gap is preserving optimal phase-matching for the basic 3WM process, while the wide second gap is used for suppressing higher-frequency mixing modes.
    The former is necessary because of a non-negligible phase-mismatch $\Delta k> 0$, which is a side-effect of the wide gap $j=2$. This mismatch corresponds to a coherence length $\xi=\pi/\Delta k$ on the order of a few hundred elementary cells, $\xi<N$, limiting the achievable signal gain. 
    However, placing the pump frequency $\omega_p$ slightly above the upper edge of the first gap makes it possible to somewhat reduce the wave number $k_p$ and, hence, compensate the mismatch Eq.~(\ref{Eq-PhaseMatching3WM}), $\Delta k\approx 0$, in a rather wide range of the signal frequency (Fig.~\ref{fig-profile}c). 
    This corresponds to a coherence length $\xi$ on the order of several thousands of elementary cells, and for $\xi \gg N$ it results in nearly exponential signal growth along the array \cite{Zorin2016}. 
    
    The wide second gap encompasses the second harmonic of the pump, $\omega_{2p}$, and covers a wide frequency range around this frequency (see Fig.~\ref{fig-profile}b). 
    The up-converted frequencies $\omega_{p+s}$ and $\omega_{p+i}$ either also fall in this gap or are located near the band edge (purple and brown dashed lines in Fig.~\ref{fig-profile}b). 
    Then either the propagation of the generated waves  is forbidden, or the phase matching for the corresponding processes, given by Eqs.~(\ref{eq-w1})-(\ref{eq-w2p}), is violated. 
    The latter condition prevents a steady amplification of the respective wave, so these tones can only grow in an undulation manner.  
	In this way, it is quite possible to effectively suppress the growth of unwanted mixing modes, while preserving the phase-matching of the basic 3WM process.

	\section{Circuit simulations}\label{section_circuitsimulations}
	\subsection{\emph{WRspice} circuit model}
	
    In order to numerically model our JTWPA circuit, we use the software \emph{WRspice} \cite{WRspice}, which is a SPICE-based circuit simulator, capable of simulating circuits containing Josephson junctions. 
    \emph{WRspice} has been used earlier to model the JTWPA without dispersion engineering of the transmission line based on rf-SQUIDs \cite{Dixon2020} and for analyzing the effect of parameter variation on the perfomance of this JTWPA \cite{OPeatain2021preprint}. 
    Compared to other approaches of analyzing a JTWPA, SPICE simulations offer more flexibility in modeling and rigorously analyzing the circuit -- an advantage which comes at the cost of higher computation power.  
    The alternative coupled-mode equations (CME) approach, often used to analyze JTWPAs \cite{Eom2012,Yaakobi2013,OBrien2014}, gives valuable insight into the physics of such devices, but relies on several assumptions and approximations, including the transition from the discrete to the continuous telegrapher's equation \cite{Yaakobi2013}, the slowly-varying amplitude approximation \cite{OBrien2014}, and the restriction to a limited number of tones included in the analysis \cite{Dixon2020}. 
    The approach of Planat et al.~\cite{Planat2020}, based on finding the solution of the discrete wave equation in the form of traveling waves, avoids most of these approximations, but does not allow including reflections of the pump wave (below we show that for our circuit the pump reflections can be quite pronounced). 	
    Moreover, it is a commonly applied approximation to omit higher orders of nonlinearity of the examined device in its theoretical description (c.f. the truncated Taylor series in Eq.~(\ref{L-vs-phi})).
    In contrast to these approaches, no such assumptions are made in a \emph{WRspice} transient analysis.  Specifically, \emph{WRspice} simulator models the conventional sinusoidal current-phase relation of the Josephson junction and thus synthesizes the current-phase relation of rf-SQUIDs in a form which Taylor expansion also contains infinite number of terms. In this way, the simulator takes into account all waves occuring due to high-order nonlinearities and their reflections.   
    
\begin{figure}[t]
		\begin{center}
			\includegraphics[width=8.6cm]{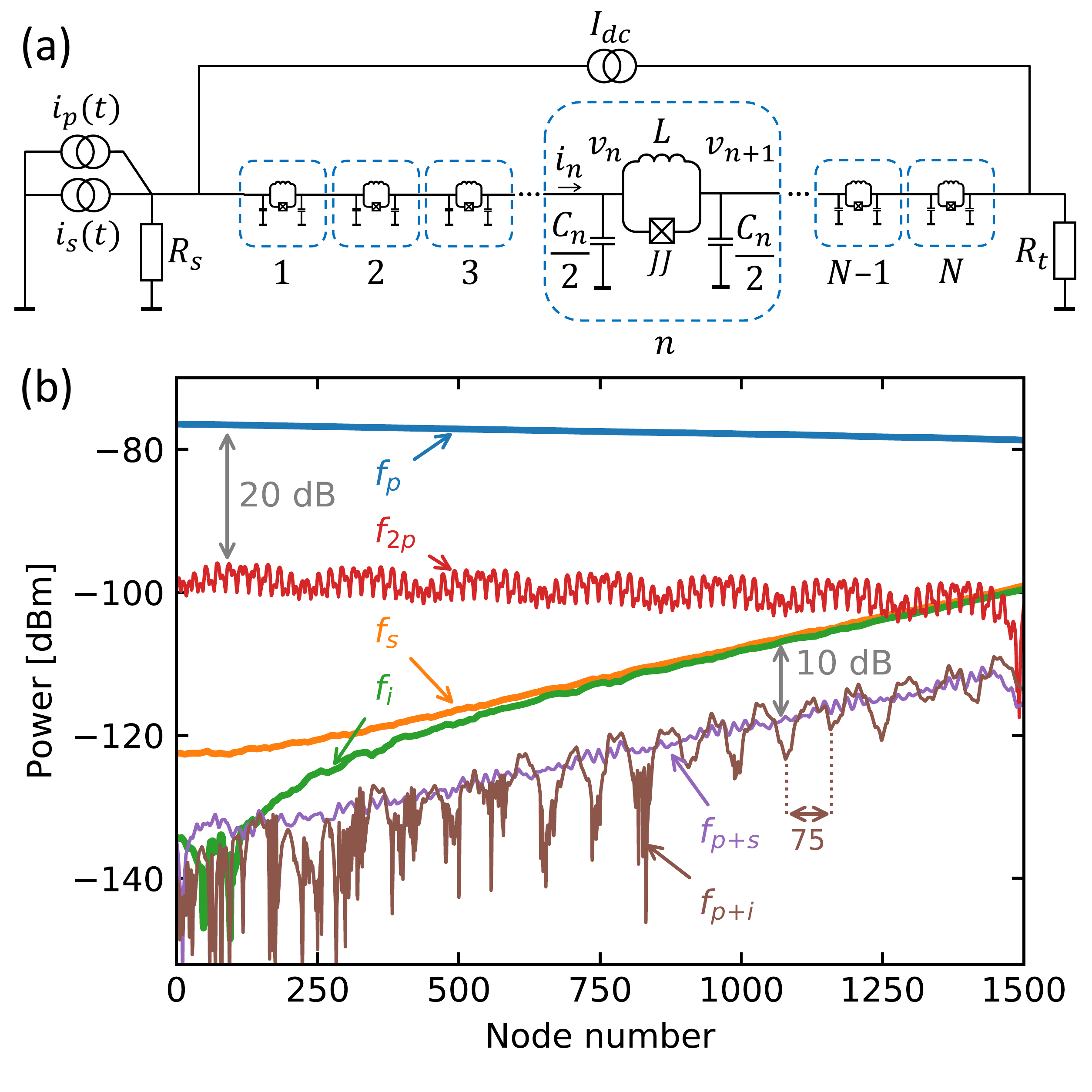}
			\caption{(a) Schematic of the JTWPA circuit  implemented in simulator \emph{WRspice}. 
			The circuit consists of $N=1500$ $\pi$-type unit cells, having capacitances $C_n$ according to the capacitance profile of Fig.~\ref{fig-profile}a, and inductances $L=84$~pH.   
			The Josephson junction (JJ) obeys a RCSJ model as described in the text,  with $I_c=1.57~\textrm{\textmu A}$, $C_J=20~\textrm{fF}$, and  $I_cR_J=16.5~\textrm{mV}$.
			Input and output of the  circuit are terminated by resistances $R_s=R_t=50$~$\Omega$, respectively. 
			Pump and signal are fed by current sources with currents $i_p(t)= \hat{I}_{p} \sin\omega_p t$ and $i_s(t)= \hat{I}_{s} \sin\omega_s t$, respectively;  a dc current $I_\textrm{dc}=9.8$~\textmu A provides the flux-biasing of all rf-SQUIDs. 
			Panel (b) shows the propagation of the six most relevant waves along the JTWPA extracted from the  \emph{WRspice} simulation. 
			The unwanted higher-frequency tones $f_{2p}$, $f_{p+s}$, and $f_{p+i}$ are significantly suppressed.
			All higher harmonics and mixing products had notably lower powers than the latter three tones. 
			The signal and idler tones rise nearly exponentially, demonstrating an almost pure 3WM process. 
			Parameters of the circuit used in this simulation are $\hat{I}_{p} = 2.0$~\textmu A, $f_{p} = 12.92$~GHz, and $f_{s} = 6.7$~GHz. 
			The signal level, $\hat{I}_{s} = 0.01$~\textmu A,  was sufficiently low, so that the amplifier saturation for the given gain of ca. 20~dB remained negligibly small. 
			}
			\label{fig-tonesvsnodes}\label{fig-wrspice}
		\end{center}
	\end{figure}

	The JTWPA circuit model and its parameters used throughout this paper are presented in Fig.~\ref{fig-wrspice}a. 
	In our model, Josephson junctions obey the RCSJ model \cite{whiteley1991, jewett1982josephson} with critical current $I_c$, junction self-capacitance $C_J$, and a linear subgap resistance $R_J$, modeling the quasiparticle losses in the Josephson junctions. The subgap resistance causes an attenuation \cite{Pozar2012} 
	\begin{equation}
	    A\;[dB]=10\,\textrm{log}_{10} (e)\, N \omega^2 L_{S0}^2/Z_0R_j, 
	\end{equation}
	which amounts to ca. 1~dB at 13~GHz. The junction gap voltage, $V_g=2.6$~mV (typical value for Nb), has no influence on the results of the simulations presented in this paper, because the Josephson junctions always remain in the superconducting state, while the microwave oscillations of the phase are small. 
	Other passive circuit components (inductances $L$ and capacitances $C_n$) are modelled as lumped and lossless components. 
	A transient analysis computes voltages $v_n(t)$ and currents $i_n(t)$ for each node $n$ as a function of time over a specified time interval. These transient quantities are then converted to the frequency domain performing Discrete Fourier Transform (see Appendix~\ref{appendix-DFT} for details) and the amplitudes and phases of the relevant tones are extracted.

	\subsection{Wave-mixing and parametric amplification}
	
	The spatial evolution of the waves of the principal tones, i.e., pump $f_p$, signal $f_s$, and idler $f_i$, plus the undesired waves with frequencies  $f_{p+s}$, $f_{p+i}$, and  $f_{2p}$, are shown in Fig.~\ref{fig-tonesvsnodes}.  For each of these tones, the powers $P_{n} =\frac{1}{2}\textrm{Re}\{ V_{n}I_{n}^{*}\}$ associated with the $n$-th node ($V_{n}$ is the complex node voltage in the frequency domain, 
	$I_n^*$ is the complex-conjugate of the complex node current $I_{n}$),  
	are plotted against the node number. 
	One can see that the signal and the idler increase almost exponentially \cite{Zorin2016},
	\begin{align}
		A_s(n) &\propto \cosh(gn)+\frac{\textrm{i}\Delta k}{2g}\sinh(gn), \label{eq-signal_vs_x} \\
		A_i(n) &\propto \sinh(gn), \label{eq-idler_vs_x}
	\end{align}
	as they propagate along the array. Here, $A_s,A_i$ are the (complex) amplitudes  of the signal and the idler, respectively, and $g$ is  an exponential gain coefficient \cite{Zorin2016}. 
	This gives evidence of parametric amplification with almost pure 3WM, remaining phase-matched all along the line, with a coherence length $\xi=2186>N$, and resulting in a signal gain of ca. 22~dB for the frequencies given in the caption of Fig.~\ref{fig-tonesvsnodes}b.  
	In contrast, the mixing process given by Eq.~(\ref{eq-w2}) is mismatched, i.e,  
	\begin{equation}
	    \Delta k_{p+i}=k_{p+i}-k_p-k_i \gg \pi/N\,,
	\end{equation}
	and thus it is incoherent, with $\xi_{p+i}=\pi/\Delta k_{p+i}=75\ll N$ (brown line). 
	The tones $f_{p+s}$ (purple line) and $f_{2p}$ (red line) are evanescent modes \cite{Erickson2017}, i.e., they fall in the wide second gap, where no propagation is possible, and show distinct beating patterns. 
	For example, the tone $f_{2p}$ appears as two predominant superimposed beating patterns with envelope periods of $\pi/(2k_p-k_m)=104$ cells each.  
	The up-conversion tones, $f_{p+s}$ and $f_{p+i}$, are each ca. 10~dB lower in amplitude than the signal, and the power of the second harmonic of the pump is about 20~dB below the pump wave power. 
	Hence these tones are strongly suppressed. 
	For example, only ca. 1\% of the pump power is converted to harmonics (all higher harmonics have even lower amplitudes than the second harmonic). 
	For comparison, in a JTWPA circuit without dispersion engineering (with otherwise identical circuit parameters) up to 70\% of the pump power is converted to harmonics (see Appendix~\ref{appendix_nogap}).

	\subsection{Gain profile} 

	\begin{figure}[t]
		\begin{center}
			\includegraphics[width=8.6cm]{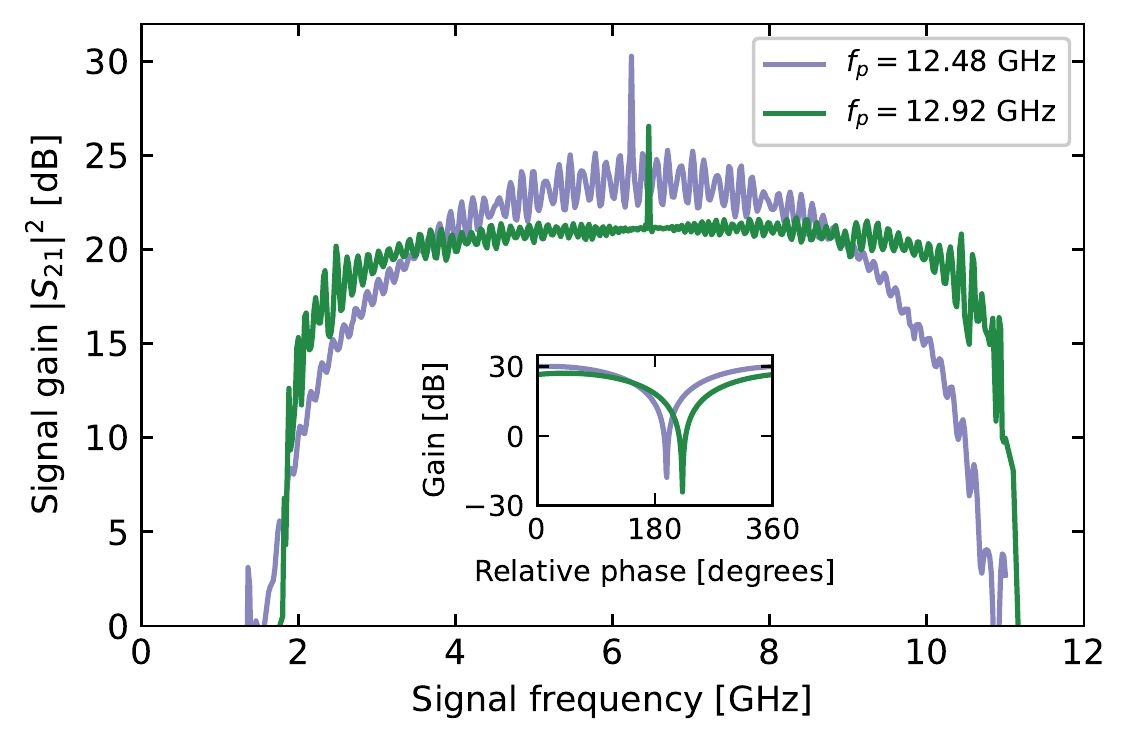}
			\caption{			Signal gain versus frequency in a JTWPA for two different pump frequencies.   
			While higher gain $|S_{21}|^2$ is obtained with $f_p=12.48~\textrm{GHz}$, the pump frequency $f_p=12.92~\textrm{GHz}$ yields better flatness, larger 3~dB-bandwidth, and greatly reduced ripple. 
			The peaks in the center of the curves occur due to degenerate 3WM, i.e., $f_s=f_i=f_p/2$. 
			In this regime, parametric amplification is phase-sensitive.
			The inset shows the phase-sensitivity of the gain at $f_s=f_p/2$, where the signal phase was tuned relative to the phase of the pump. The phase-sensitive gain varies periodically between amplification and deamplification.
			In the case of deamplification the gain profile would show a dip in its center. 
			The maximum degenerate gain is ca. 6~dB higher than the  nondegenerate gain in the vicinity of $f_p/2$, while the maximum extinction ratio is approaching 48~dB for $f_p=12.48~\textrm{GHz}$ and 51~dB for $f_p=12.92~\textrm{GHz}$. 
			The pump amplitude was $\hat{I}_{p} = 1.8$~\textmu A in these simulations. 
			}
			\label{fig-dge-bw+backamp}
		\end{center}
	\end{figure}

	Continuous broadband gain was found only for pump frequencies slightly above the first gap. 
	Pump frequencies in a range of roughly 12.4--13.2~GHz permit sufficient phase matching for broadband signal gain of $G=|S_{21}|^2>20$~dB at $\hat{I}_p=1.8$~\textmu A (incident pump power $P_{p,in}\approx-78$~dBm). 
	Varying the pump frequency alters $\Delta k(f_s)$ (see Fig.~\ref{fig-profile}c) and thus influences the shape of the gain profile. 
	The dependence of the gain on the signal frequency is shown in Fig.~\ref{fig-dge-bw+backamp} for two different pump frequencies.  
	The gain profile for $f_p=12.48$~GHz shows a nearly elliptical shape, since the phases of pump, signal, and idler are matched best in the center of the signal band, $f_s=f_p/2$. 
	For $f_p=12.92$~GHz, phase-matching is imperfect in the center, but perfect ($\Delta k=0$) farther apart from the center (see the orange curve in Fig.~\ref{fig-profile}c). This leads to a slight drop in the center of the gain profile, when compared to that for $f_p=12.48$~GHz, but also to better flatness and larger 3~dB-bandwidth of ca. 7.2~GHz~$\approx0.56\,f_p$. 
    The resulting bandwidth is furthermore a function of the convexity of the dispersion relation \cite{Malnou2021}, which is defined by the width of the gaps or, equivalently, the modulation depth of the capacitance profile $C_n$. 
    A smaller modulation depth leads to a wider range of phase-matched signal frequencies, but also reduces the phase-mismatch for unwanted processes. Thus the modulation depth is a tradeoff between bandwidth and the suppression of unwanted parametric processes. 
    
    The distinct peaks at $f_s=6.24$~GHz and 6.46~GHz for $f_p=12.48$~GHz and 12.92~GHz, respectively, emerge due to degenerate 3WM, where the signal and idler frequencies coincide,  $f_s=f_i=f_p/2$. 
    As long as the phase of the idler is rigidly connected to the phases of the pump and the signal (see, e.g., Eq.~(19) in Ref.~\cite{Zorin2016}), the signal and the idler interfere either constructively (leading to amplification) or destructively (leading to deamplification), depending on the phase of the signal relative to that of the pump. 
    Respective simulations are shown in the inset in Fig.~\ref{fig-dge-bw+backamp}. 
    This interesting regime, where single-mode squeezing is possible \cite{Perelshtein21}, is easily accessible in a 3WM-JTWPA, since the pump frequency is well separated from the signal and the idler, $f_p=2f_{s,i}$ (in contrast to a conventional 4WM-JTWPA, where the signal and the idler coincide with the strong pump tone, $f_p=f_{s,i}$, that masks the effect). As was recently demonstrated by Qiu et al. \cite{Qiu2022preprint}, phase-sensitive amplification is, however, possible using two pump waves ($f_{p1} < f_s = f_i < f_{p2}$) in a 4WM-JTWPA, and they report a a large extinction ratio of 56~dB. 
    
\begin{figure}[b]
		\begin{center}
			\includegraphics[width=8.6cm]{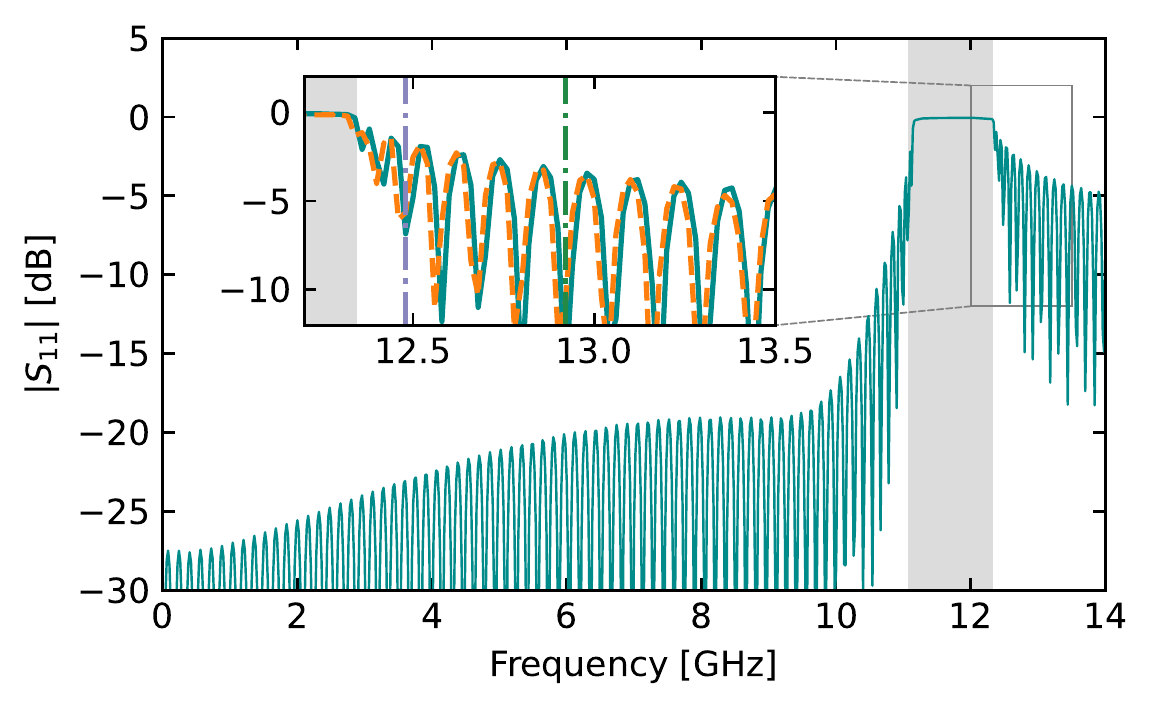}
			\caption{The reflection coefficient $|S_{11}|$ for typical frequencies of signal, idler, and pump. 
			The quasi-periodic patterns are the side-lobes of the first stop-band (grey-shaded area).   
			The inset shows a close-up view of $|S_{11}|$ in the pump frequency range. 
			As the pump power increases ($\hat{I}_p = 0.01$~\textmu A for the cyan curves and $\hat{I}_p = 1.8$~\textmu A for the dashed orange curve), the gap and its side-lobes shift slightly to lower frequencies. 
			To minimize pump reflection, the pump frequency is placed in a minimum between two adjacent side-lobes; violet and green vertical lines mark the pump frequencies used in Fig.~\ref{fig-dge-bw+backamp}. 
			The curves are obtained by \emph{WRspice} simulations where only one sine-wave (pump) current was injected ($\hat{I}_s = 0$). 
			}
			\label{fig-side_lobes}
		\end{center}
	\end{figure}

    The ripple in the gain profile (Fig.~\ref{fig-dge-bw+backamp}) is an unwanted but expected feature for any JTWPA with a slight impedance mismatch $Z\ne Z_0$ ($Z_0$ is the impedance of the external circuit connected to the JTWPA)  \cite{Eom2012,Planat2020,Zhao2021}. 
    For the curves in Fig.~\ref{fig-dge-bw+backamp} the average value $\overline{Z}= 52~\Omega$  (valid for low frequencies, i.e., wavelengths $\gg m$) is close to $Z_0$. 
    The impedance mismatch is in this case mainly caused by the frequency dependence of $Z$ due to relatively low frequencies $f_J$ and $f_0$ and significant dispersion in the range from 0 to $f_p$ engineered by the periodic loadings.  
    This impedance mismatch causes signal reflections at both ends of the array, and the back- and forth-propagating signal wave creates a Fabry-Perot-like interference pattern \cite{Planat2020}. 
    The frequency-spacing in this pattern corresponds to twice the electrical length of the array, $\Delta f=v_p/2\ell\approx\omega_0/2N=160$~MHz, where $v_p$ is the average phase velocity. 
    The idler wave is also multiply reflected, but travels at a slightly different phase velocity due to small chromatic dispersion. 
    The parametric interaction of signal and idler between their multiple reflections leads to ripple having two superimposed, slightly different, periods. 
    Furthermore, it was observed in simulations (not shown)
    that the amount of ripple increases with increasing gain, which is a fingerprint of Fabry-Perot-like interferences in JTWPAs due to the amplification the signal wave experiences as it travels between its multiple reflections \cite{Planat2020}.
    In the same way, the occurrence of backward amplification, i.e., the 3WM interaction of the reflected signal (wave-number $-k_s$) with the reflected pump ($-k_p$)  \cite{Zhao2021}, can further enhance ripple. 
    For pump frequencies closer to the gap the reflectivity and hence the backward gain are high (see inset in Fig.~\ref{fig-side_lobes}), which accounts for the higher amount of ripple in the gain profile of $f_p=12.48$~GHz compared to that of $f_p=12.92$~GHz (Fig.~\ref{fig-dge-bw+backamp}).

    To decrease ripple and prevent strong backward-traveling signal waves, the reflectivity 
    at the output and the input of the array should be minimized for both the signal, say, between 1 and 11~GHz, and the pump frequency. 
    Therefore, sufficiently good impedance matching should be achieved. 
    The passive circuit components $L$ and $C_n$, designed to meet the target value  $Z=Z_0$ 
    at the envisaged working point $\phi_\textrm{dc}^\textrm{opt}(\phi_e)\approx \pi/2+\beta_L$, 
    can only be manufactured within certain technological margins. 
    However, the dependence $Z(\phi_\textrm{dc})$ (Eq.~(\ref{lineimpedance})) is rather steep in the vicinity of that working point,	such that small deviations of $L$ and $C_n$ from their target values can be compensated by fine-tuning of $\phi_e$ without degrading the amplifier performance dramatically. 
	For example, $\Delta Z=\pm$10\% is achieved by  $\Delta\phi_e\approx\pm0.4$~rad. The corresponding change of $\beta$ is smaller than 20\%, which can be compensated by changing the pump amplitude. 
	The Kerr coefficient, $|\gamma|<0.08$, is still only a fraction of $|\beta|$, and the resulting phase-mismatch due to SPM and XPM can be compensated by re-adjusting the pump frequency anyway. 
	Therefore, the circuit design allows \emph{in-situ} fine tuning of the transmission line impedance by an external magnetic field and thereby makes it possible to reduce unwanted reflections of microwaves at the output and the input.

    It is not possible, however, to fully suppress gain ripple  by fine-tuning of the impedance. 
    Due to the periodic variation of the ground capacitances, the impedance $Z$, with average value $\overline{Z}$ (valid only for wavelengths $\gg m$), shows a slight periodical variation with frequency. 
    This variation appears as side-lobes of the gaps (being the main-lobes) in the reflection coefficient $S_{11}(f)$, which is presented in Fig.~\ref{fig-side_lobes}. 
    The sidelobes have a periodicity of $\Delta f\approx\omega_0/2N=160$~MHz, similar to that of the gain ripple.  
    For frequencies closer to the gap this period is reduced. 
    The gaps and their side-lobes shift slightly to smaller frequencies as the pump power is increased (inset in Fig.~\ref{fig-side_lobes}), which is due to a small residual Kerr nonlinearity \cite{Planat2020}. 
    This shift is small, however, when compared to that reported in Ref.~\cite{Planat2020}. 
    To minimize pump reflections, the pump frequency is placed in the mimimum between two adjacent side-lobes. 
    In the signal band (ca. 3...9~GHz), sufficiently far below the first gap, the reflection coefficient of up to $-19$~dB is rather small, but still leads to weak multireflections, and thus to notable gain ripple (Fig.~\ref{fig-dge-bw+backamp}).
    When the average impedance is matched, $\overline{Z}=Z_0$, these side-lobes are the main cause of ripple in the gain profile. 
    It should be noted that the occurence of side-lobes is immanent to the concept of periodic loadings and cannot be resolved by standard impedance matching techniques easily. 
    However, its effect can be possibly mitigated using the apodization technique, a well-established method for side-lobe suppression in optical fibre Bragg gratings \cite{Southwell1989,AbuSafia1993}. 
    Adapting this technique to our concept should be possible by designing a non-uniform modulation depth of the periodic loadings (instead of a uniform modulation depth as in the present paper),  enveloped by a suitable apodization function, e.g., a truncated Gaussian.
    Potentially this could be a method to strongly reduce unwanted reflections  and ripple in the gain profile.

	\section{Conclusion}

        In conclusion, we have proposed a modified design of a lumped-element Josepshon traveling-wave parametric amplifier using periodic loadings with the shape shown in Fig.~\ref{fig-profile}a and illustrated by circuit simulations the possibility to achieve gain of 20 dB in the frequency range from 3 GHz to 9 GHz. 
        We presented a simulation approach which includes all occuring tones, reflections from the amplifier terminations, etc.,  and has no assumptions on the order of nonlinearities produced by the Josephson junctions. 
        Our implementation of periodic loadings enables both reasonable phase matching of the basic 3WM process and effective suppression of unwanted high-frequency modes. 
        The design of the device remains simple and close to the original design \cite{Zorin2016}, i.e., it does not need any additional elements in the amplifier architecture. 
        Furthermore, it is possible to slightly tune the line impedance of the device by a magnetic flux bias. 
        We believe that further improvement of the JTWPA characteristics is possible by applying periodic loadings with a more sophisticated design, including periodic variation of the rf-SQUID parameters. This could be done, for example, by varying the sizes of the Josephson junctions along the array according to the recently proposed technique of the Floquet-mode JTWPA \cite{Peng2022}.

        We think that periodic loadings in JTWPAs with 3WM may open the way to practical low-noise parametric devices with sufficiently large bandwidth, which are a key enabling technology for quantum communication and quantum computing circuits. 
        Due to good phase matching, a JTWPA of this type presents a circuit with remarkable properties, which may allow, for example, creating various quantum states of microwaves in a wide frequency range \cite{Greco2021, Fasolo2021}.

	\begin{acknowledgments}

		The authors would like to thank Lukas Grünhaupt for useful discussions and Tom Dixon, Dominik Müller and Victor Rogalya for their assistance in setting up the simulator platform.		
		This work has received funding from the EMPIR programme
        (project ParaWave 17FUN10) co-financed by the Participating
        States and from the European Union’s Horizon 2020
        research and innovation programme.
        This work was also supported
        by the German Federal Ministry of Education and Research (BMBF)
        within the framework programme “Quantum technologies –
        from basic research to market” (Grant No. 13N15949).
		C.K. gratefully acknowledges the support of the Braunschweig International Graduate School of Metrology B-IGSM and the DFG Research Training Group 1952 Metrology for Complex Nanosystems.

	\end{acknowledgments}

	\section*{Author declarations}
	\subsection*{Conflict of Interest}
	The authors have no conflicts to disclose.
		
	\subsection*{Author Contributions}
	\textbf{V. Gaydamachenko} and \textbf{C. Kissling} contributed equally to this work. 

    \textbf{V. Gaydamachenko}: Investigation (equal); methodology (equal); software (equal); validation (equal); visualisation (equal); writing - original draft (equal).
    \textbf{C. Kissling}: Investigation (equal); formal analysis (lead); methodology (equal); software (equal); validation (equal); visualisation (equal); writing - original draft (equal).
    \textbf{R. Dolata}: Funding acquisition (lead); Project administration (lead); supervision (equal); writing - review \& editing (supporting).
    \textbf{A. B. Zorin}: Conceptualization (lead); methodology (supporting); supervision (equal); writing - original draft (equal); writing - review \& editing (lead).
	
	\section*{Data Availability Statement}

    The data that support the findings of this study are openly available in Zenodo repository at https://doi.org/10.5281/zenodo.7092947.

	\appendix

    \section{JTWPA  without dispersion engineering}\label{appendix_nogap}

    \begin{figure}[b]
		\begin{center}
			\includegraphics[width=8.6cm]{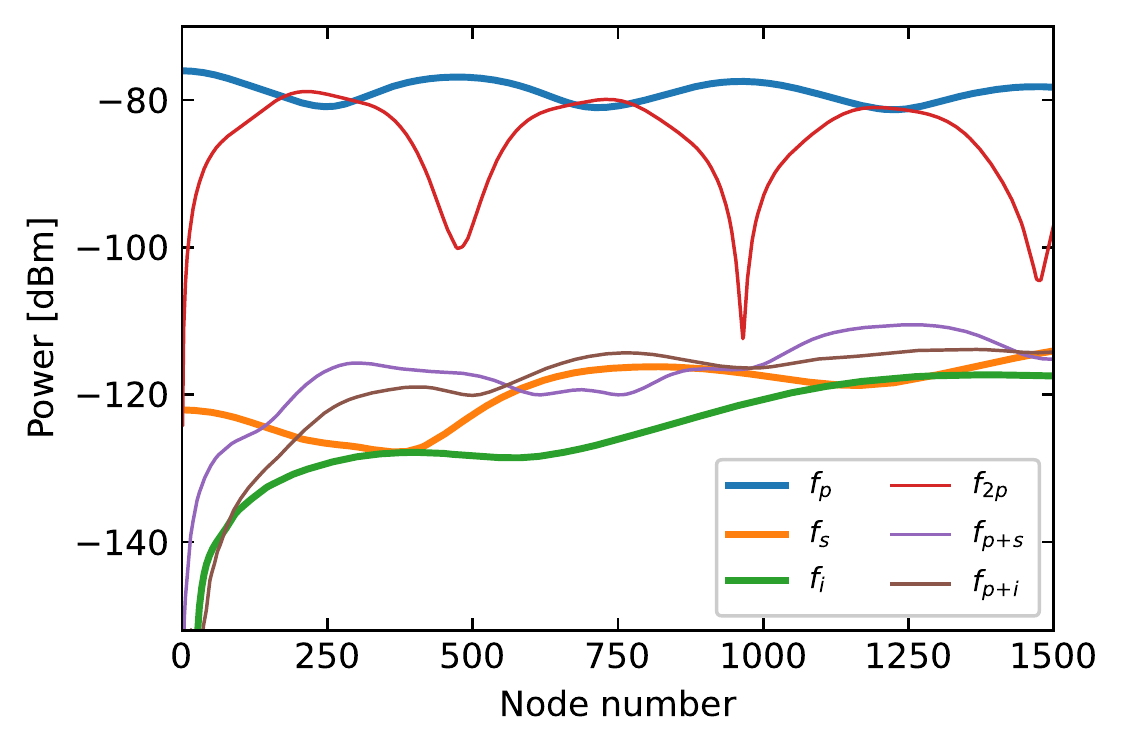}
			\caption{The dynamics of the major six waves propagating along the homogeneous JTWPA without dispersion engineering, i.e., having identical ground capacitors throughout the array. 
			Power is transferred alternatingly between the pump and the second harmonic of the pump; and the pump is decreased in its mimima by 5~dB, where the second harmonic power is larger  than the pump power. 
			The signal and idler waves cannot grow monotonically due to their strong interaction with the up-conversion tones $f_{p+s}$ and $f_{p+i}$. The signal gain is 8~dB. 
			Parameters used in this simulation are $\hat{I}_{p} = 2.0$~\textmu A, $f_{p} = 12.92$~GHz and $f_{s} = 8.0$~GHz, and the circuit parameters are as described in the text. 
			}
			\label{fig-powers_nogap}
		\end{center}
	\end{figure}
    
    In a JTWPA without dispersion engineering, i.e., with a homogeneous transmission line, the signal gain is limited by power leakage from the main tones to unwanted tones. 
    To illustrate the benefit of our proposed dispersion engineering approach,  we present here the circuit simulations of a  homogeneous JTWPA circuit for comparison with Fig~\ref{fig-tonesvsnodes}b and Fig.~\ref{fig-dge-bw+backamp}. 
    For comparability, the  circuit parameters were chosen identical to those of our proposed dispersion engineered JTWPA,
    apart from the homogeneous capacitance profile with constant ground capacitance values $C_n=\overline{C}=40$~fF for all $n=1...N$. The resulting characteristic frequency $\omega_0$ was identical to that of our proposed JTWPA circuit. 
    The dispersion of this circuit, with the dispersion relation given by Eq.~(\ref{k-vs-low-w}), is low in the frequency range $0...f_{2p}$. 
    Therefore, the phase mismatch of the second harmonic, $\Delta k_{2p}=k_{2p}-2k_p$, is small, and the  coherence length of the second harmonic, $\xi_{2p}=\pi/\Delta k_{2p}$, is on the order of 500 cells (see Fig.~\ref{fig-powers_nogap}). 
    This allows the second harmonic to grow to high amplitudes on the order of the pump amplitude, before the direction of power transfer reverses and power flows back to the pump wave. 
    This power leakage to the second harmonic  depletes the pump periodically to powers lower than 30\% of its incident value $P_{p,in}$. Consequently, the average pump amplitude, needed for efficient parametric amplification, is significantly reduced. 
    
\begin{figure}[t]
		\begin{center}
			\includegraphics[width=8.6cm]{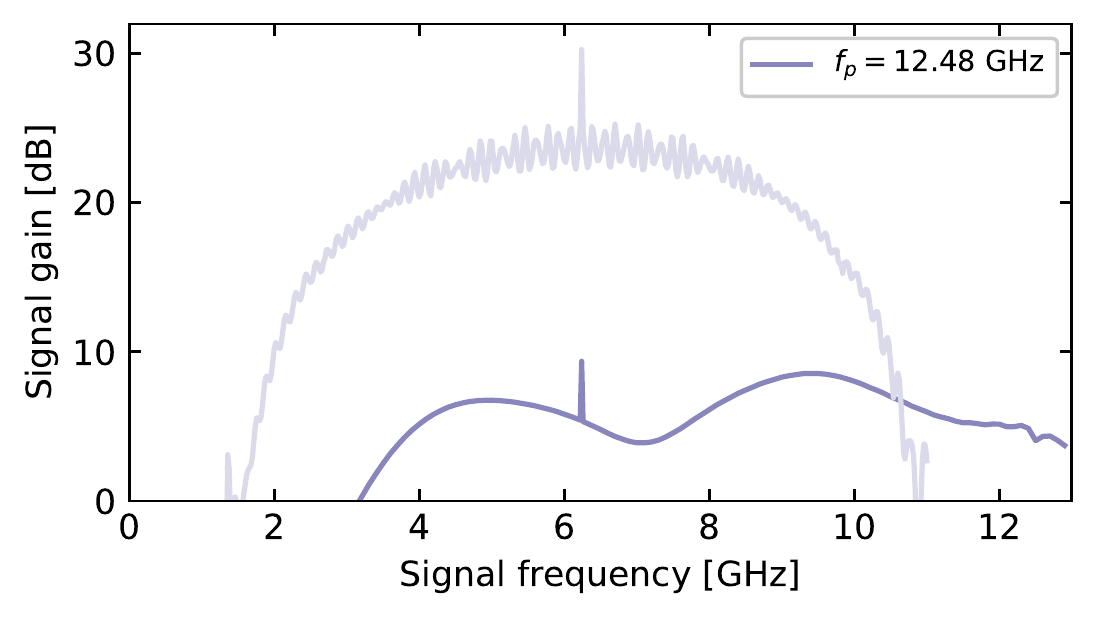}
			\caption{Signal gain versus frequency in the JTWPA without dispersion engineering (dark-purple curve).  For comparison the respective curve of a dispersion-engineered JTWPA (light-purple curve, c.f. Fig.~\ref{fig-dge-bw+backamp}) is shown. 
			Without dispersion engineering the signal gain is limited to ca. 9~dB and the gain profile is asymmetric. The distinctive peaks stem from the phase-sensitive amplification at the degenerate 3WM, $f_s=f_p/2$. 
			The pump amplitude is $\hat{I}_{p} = 1.8$~\textmu A and the circuit parameters are as described in the text.  
			}
			\label{fig-gain_nogap}
		\end{center}
	\end{figure}

    Similar reasons allow a strong power transfer from the signal and the idler tones to the up-conversion tones $f_{p+s}$ and $f_{p+i}$, whose amplitudes are either in the order of or even higher than the signal and idler amplitudes, respectively.  
    These tones decimate the signal and the idler, so that an exponential growth of the signal is impossible and the gain is limited to rather small values of less than 10 dB (Fig.~\ref{fig-gain_nogap}, compare with Fig.~8 in Ref.~\cite{Dixon2020}). 
    Note that the signal wave shown by the orange line in Fig.~\ref{fig-powers_nogap} is less affected by up-conversion than the idler wave (green line). 
    This is due to  worse phase matching (shorter coherence length $\xi_{p+s}=735$) than than that of its counterpart ($\xi_{p+i}=1402$) because of stronger dispersion at higher frequencies (Eq.~(\ref{k-vs-low-w})). This explains the assymetry of the gain profile in Fig.~\ref{fig-gain_nogap}.

	\section{Transfer-matrix method} \label{appendix-TMM} 
	
	The transfer-matrix method is a standard approach in optics \cite{Mackay2020} and microwave engineering \cite{Pozar2012} and is particularly useful for linear circuits consisting of cascaded two-ports. Here we apply this method to the JTWPA which is a cascade of $N$ elementary two-ports shown in Fig.~\ref{fig-EqvSchm}b. Each two-port $n$ is described by a transfer matrix $T_n$ (also referred to as ABCD-matrix) 
	\begin{equation}
		T_n=\left(
		\begin{matrix}
			A_n & B_n  \\
			C_n & D_n 
		\end{matrix}
		\right) ,
		\label{ABCD-matrix}
	\end{equation}
	with the coefficients
	\begin{align}
        A_n &= 1-\frac{1}{2}\frac{\omega^2 L_{S0} C_{n}}{1-\omega^2 C_J L_{S0}}, \\
        B_n&= \frac{\textrm{i}\omega L_{S0}}{1-\omega^2 C_J L_{S0}}, \\
		C_n &= \textrm{i}\omega C_n-\frac{1}{4}\frac{\textrm{i}\omega^3 L_{S0} C_n^2 }{1-\omega^2 C_J L_{S0}}, \\
		D_n&= A_n
        \label{ABCD-coefficients}
    \end{align}
	These $N$ transfer matrices $T_n$ are then cascaded to a system transfer matrix
	\begin{equation}
		T_N=\left(
		\begin{matrix}
			A & B  \\
			C & D 
		\end{matrix}
		\right) = \left(T_{01}^\kappa\,T_{02}^\mu\,T_{01}^\kappa\,T_{03}^\nu\right)^\frac{N}{m},
		\label{ABCD-cascaded}
	\end{equation}
	where the indices $01,02$, and $03$ refer to the respective ground capacitor $C_n$  (see Fig.~\ref{fig-profile}a), the exponents $\kappa,\mu$ and $\nu$ are integers and denote the number of elementary two-ports of each variant (e.g., $\kappa=\mu=\nu=5$ in  our circuit) in one section, and in total there are $N/m$ such cascaded sections, with $m=2\kappa+\mu+\nu$. The system transfer matrix can then be transformed to a scattering matrix \cite{Pozar2012} for a given reference impedance $Z_0$. The input reflection $S_{11}$ and forward transmission $S_{21}$ (for this linearized circuit $S_{12}=S_{21}$ and $S_{22}\approx S_{11}$) are given by the following expressions:
	\begin{equation}
		S_{11}=\frac{A+B\,/\,Z_0-C\,Z_0-D}{A+B\,/\,Z_0+C\,Z_0+D},
		\label{S11_from_TMM1}
	\end{equation}
	\begin{equation}
		S_{21}=\frac{2}{A+B\,/\,Z_0+C\,Z_0+D}.
		\label{S11_from_TMM2}
	\end{equation}

	\section{Simulation setup and post processing}\label{appendix-signal_gain}\label{appendix-S-parameters}\label{appendix-DFT}

    All simulations presented in this paper were performed by the open-source software \emph{WRspice} as transient analysis. 
    A \emph{WRspice} netlist containing our JTWPA circuit can be found in Ref.~\cite{Zenodo-rep}. 
    As a default initial condition, internal routines of \emph{WRspice} set all currents through inductors and all voltages across Josephson junctions to zero. In this way, the flux-phase relations in loops containing Josephson junctions are imposed \cite{WRspice}. 
    The dc bias current source is ramped up from zero to its final value of $9.8$~\textmu A within $0.4$~ns to avoid large transients. 
    AC sources are active from the beginning of simulation.
    The transient response of the perturbation caused by the sources propagates and takes ca. $t_{tr}=\ell/v = N/\sqrt{L_{S0}\overline{C}} = 3.13$~ns to reach the end of the circuit. 
    The circuit is close to its steady state after 10~ns. 
    Time domain data is saved between 10~ns and 60~ns, giving a total sampling time $T= 50$~ns, which defines the resolution of the Fourier transform $\Delta f = 1/T = 20$~MHz. 
    Increasing the delay time of 10~ns does not change the results notably. 
    The maximum time step of transient analysis is set to $\Delta t=4.0$~ps. Note, that the maximum time step is chosen so small that the adaptive time step control of \emph{WRspice} (automatically adapting the time step to optimize simulation time and accuracy) is overridden. Thus, the resulting time series is equidistant and $\Delta t$ defines the maximum frequency of the Fourier transform, $f_{max} = 1/2\Delta t = 125$~GHz $> f_{J}, f_{0}$. 

    In the post processing the output of \emph{WRspice}, the time-series of voltages $v_n(t)$ and currents $i_n(t)$ for each individual node $n$, is transformed to the frequency domain by applying a Discrete Fourier Transform (DFT). 
    In our implementation the DFT is defined as
	\begin{equation}
		I_{q,n} =\frac{1}{M} \sum_{\tau=0}^{M-1} i_{n} \exp\bigg(- \frac{2\pi \textrm{i}}{M}\tau q \bigg),\quad q = 0, ... , M-1  ,
		\label{eq-I-DFT}
	\end{equation}
	\begin{equation}
		V_{q,n} =\frac{1}{M} \sum_{\tau=0}^{M-1} v_{n} \exp\bigg(- \frac{2\pi \textrm{i}}{M}\tau q \bigg),\quad q = 0, ... , M-1  .
		\label{eq-V-DFT}
	\end{equation}
    Here $M$ is the size of the time-series and $q$ is the index defining the corresponding frequencies $f_{q} = q\Delta f$. 
    After DFT, we extract currents $I_n(f=f_\textrm{j})=I_{\textrm{j},n}$ and voltages $V_n(f=f_\textrm{j})=V_{\textrm{j},n}$ for each tone with index $\textrm{j}\in \{s,\,i,\,p,\,p\!+\!s,\,p\!+\!i,\,2p\}$ and for each individual node $n$. We have chosen $T$ such that all $f_\textrm{j}$ coincide with a frequency sample $\tilde q\Delta f$, where $\tilde q \in \{0, ... , M\!-\!1\}$.  
    
	Afterwards, the scattering parameters $S_{11}$ and $S_{21}$ can be generically calculated for a given frequency as follows:
	\begin{equation}
		S_{11}=\frac{V_{in}-Z_{0}I_{in}}{V_{in}+Z_{0}I_{in}}=\frac{Z_{in}-Z_{0}}{Z_{in}+Z_{0}}=\Gamma_{in},
		\label{S11-def}
	\end{equation}
	\begin{equation}
		S_{21}=\frac{V_{out}+Z_{0}I_{out}}{V_{in}+Z_{0}I_{in}}=\frac{2V_{out}}{V_{in}+Z_{0}I_{in}},
		\label{S21-def}
	\end{equation}
	where we used the relation $I_{out}=V_{out}/Z_{0}$ and the value of the input impedance $Z_{in}=V_{in}/I_{in}$; $\Gamma_{in}$ is the input reflection coefficient, and subscripts $in$ and $out$ denote the input and output quantities of the JTWPA circuit, i.e., the quantities at the first and last node of the circuit, $n=1$ and $n=N+1$, respectively. 
	In the simulations performed in this paper, only the forward transmission  $S_{21}$ and the input reflectivity $S_{11}$ are determined, because the current sources driving the signal and pump currents are connected to the JTWPA input only (Fig.~\ref{fig-wrspice}a). 
	This resembles the practical scenario of signal and pump sources connected to the input of the JTWPA, and a passive load (circulators, filters, further amplifier stages) at the output of the JTWPA. It is worth mentioning, that in the presence of a strong pump wave a JTWPA is a non-reciprocal device, i.e., the forward and backward transmissions of the signal are not identical, $S_{21}\ne S_{12}$. 
	In the ideal case the signal is amplified only when travelling in forward but not in backward direction. 
	
	Typically an amplifier like a JTWPA is embedded in an experimental environment having a characteristic impedance of $Z_{0}=50\Omega$, which we chose as a reference impedance. 
	All \emph{WRspice} simulations performed in this research use a source ($Z_S$) and load ($Z_L$) impedance $Z_S=Z_L=Z_{0}$. 
	However, the characertistic impedance $Z$ of the JTWPA is generally not purely real-valued but possesses a non-vanishing imaginary part, being a side effect of the engineered periodic loadings. 
	Due to the slight impedance mismatch, $Z\ne Z_{0}$, there is some ambiguity concerning the definition of two-port power gain of the amplifier. To account for reflections of the signal wave at the amplifier's input and output, the definition of the transducer power gain \cite{Pozar2012} is used, which, in the case of $Z_S=Z_L=Z_{0}$, is
	\begin{equation}
		G=P_L/P_{A}=|S_{21}|^2,
		\label{G-transducergain}
	\end{equation}
	where $P_L$ and $P_{A}$ denote the power at the load and the power available from the source, respectively. 
    
	\bibliography{VGaydamacheko_CKissling_TWPA}

\providecommand{\noopsort}[1]{}\providecommand{\singleletter}[1]{#1}%
\begin{thebibliography}{53}%
\makeatletter
\providecommand \@ifxundefined [1]{%
 \@ifx{#1\undefined}
}%
\providecommand \@ifnum [1]{%
 \ifnum #1\expandafter \@firstoftwo
 \else \expandafter \@secondoftwo
 \fi
}%
\providecommand \@ifx [1]{%
 \ifx #1\expandafter \@firstoftwo
 \else \expandafter \@secondoftwo
 \fi
}%
\providecommand \natexlab [1]{#1}%
\providecommand \enquote  [1]{``#1''}%
\providecommand \bibnamefont  [1]{#1}%
\providecommand \bibfnamefont [1]{#1}%
\providecommand \citenamefont [1]{#1}%
\providecommand \href@noop [0]{\@secondoftwo}%
\providecommand \href [0]{\begingroup \@sanitize@url \@href}%
\providecommand \@href[1]{\@@startlink{#1}\@@href}%
\providecommand \@@href[1]{\endgroup#1\@@endlink}%
\providecommand \@sanitize@url [0]{\catcode `\\12\catcode `\$12\catcode
  `\&12\catcode `\#12\catcode `\^12\catcode `\_12\catcode `\%12\relax}%
\providecommand \@@startlink[1]{}%
\providecommand \@@endlink[0]{}%
\providecommand \url  [0]{\begingroup\@sanitize@url \@url }%
\providecommand \@url [1]{\endgroup\@href {#1}{\urlprefix }}%
\providecommand \urlprefix  [0]{URL }%
\providecommand \Eprint [0]{\href }%
\providecommand \doibase [0]{https://doi.org/}%
\providecommand \selectlanguage [0]{\@gobble}%
\providecommand \bibinfo  [0]{\@secondoftwo}%
\providecommand \bibfield  [0]{\@secondoftwo}%
\providecommand \translation [1]{[#1]}%
\providecommand \BibitemOpen [0]{}%
\providecommand \bibitemStop [0]{}%
\providecommand \bibitemNoStop [0]{.\EOS\space}%
\providecommand \EOS [0]{\spacefactor3000\relax}%
\providecommand \BibitemShut  [1]{\csname bibitem#1\endcsname}%
\let\auto@bib@innerbib\@empty
\bibitem [{\citenamefont {Movshovich}\ \emph {et~al.}(1990)\citenamefont
  {Movshovich}, \citenamefont {Yurke}, \citenamefont {Kaminsky}, \citenamefont
  {Smith}, \citenamefont {Silver}, \citenamefont {Simon},\ and\ \citenamefont
  {Schneider}}]{Movshovich1990}%
  \BibitemOpen
  \bibfield  {author} {\bibinfo {author} {\bibfnamefont {R.}~\bibnamefont
  {Movshovich}}, \bibinfo {author} {\bibfnamefont {B.}~\bibnamefont {Yurke}},
  \bibinfo {author} {\bibfnamefont {P.~G.}\ \bibnamefont {Kaminsky}}, \bibinfo
  {author} {\bibfnamefont {A.~D.}\ \bibnamefont {Smith}}, \bibinfo {author}
  {\bibfnamefont {A.~H.}\ \bibnamefont {Silver}}, \bibinfo {author}
  {\bibfnamefont {R.~W.}\ \bibnamefont {Simon}},\ and\ \bibinfo {author}
  {\bibfnamefont {M.~V.}\ \bibnamefont {Schneider}},\ }\bibfield  {title}
  {\enquote {\bibinfo {title} {Observation of zero-point noise squeezing via a
  {Josephson}-parametric amplifier},}\ }\href
  {https://doi.org/10.1103/PhysRevLett.65.1419} {\bibfield  {journal} {\bibinfo
   {journal} {Phys. Rev. Lett.}\ }\textbf {\bibinfo {volume} {65}},\ \bibinfo
  {pages} {1419--1422} (\bibinfo {year} {1990})}\BibitemShut {NoStop}%
\bibitem [{\citenamefont {Castellanos-Beltran}\ \emph
  {et~al.}(2008)\citenamefont {Castellanos-Beltran}, \citenamefont {Irwin},
  \citenamefont {Hilton}, \citenamefont {Vale},\ and\ \citenamefont
  {Lehnert}}]{Castellanos-Beltran2008}%
  \BibitemOpen
  \bibfield  {author} {\bibinfo {author} {\bibfnamefont {M.~A.}\ \bibnamefont
  {Castellanos-Beltran}}, \bibinfo {author} {\bibfnamefont {K.~D.}\
  \bibnamefont {Irwin}}, \bibinfo {author} {\bibfnamefont {G.~C.}\ \bibnamefont
  {Hilton}}, \bibinfo {author} {\bibfnamefont {L.~R.}\ \bibnamefont {Vale}},\
  and\ \bibinfo {author} {\bibfnamefont {K.~W.}\ \bibnamefont {Lehnert}},\
  }\bibfield  {title} {\enquote {\bibinfo {title} {Amplification and squeezing
  of quantum noise with a tunable {Josephson} metamaterial},}\ }\href
  {https://doi.org/10.1038/nphys1090} {\bibfield  {journal} {\bibinfo
  {journal} {Nat. Phys.}\ }\textbf {\bibinfo {volume} {4}},\ \bibinfo {pages}
  {929--931} (\bibinfo {year} {2008})}\BibitemShut {NoStop}%
\bibitem [{\citenamefont {Hatridge}\ \emph {et~al.}(2011)\citenamefont
  {Hatridge}, \citenamefont {Vijay}, \citenamefont {Slichter}, \citenamefont
  {Clarke},\ and\ \citenamefont {Siddiqi}}]{Hatridge2011}%
  \BibitemOpen
  \bibfield  {author} {\bibinfo {author} {\bibfnamefont {M.}~\bibnamefont
  {Hatridge}}, \bibinfo {author} {\bibfnamefont {R.}~\bibnamefont {Vijay}},
  \bibinfo {author} {\bibfnamefont {D.~H.}\ \bibnamefont {Slichter}}, \bibinfo
  {author} {\bibfnamefont {J.}~\bibnamefont {Clarke}},\ and\ \bibinfo {author}
  {\bibfnamefont {I.}~\bibnamefont {Siddiqi}},\ }\bibfield  {title} {\enquote
  {\bibinfo {title} {Dispersive magnetometry with a quantum limited {SQUID}
  parametric amplifier},}\ }\href {https://doi.org/10.1103/PhysRevB.83.134501}
  {\bibfield  {journal} {\bibinfo  {journal} {Phys. Rev. B}\ }\textbf {\bibinfo
  {volume} {83}},\ \bibinfo {pages} {134501} (\bibinfo {year}
  {2011})}\BibitemShut {NoStop}%
\bibitem [{\citenamefont {Vijay}, \citenamefont {Slichter},\ and\ \citenamefont
  {Siddiqi}(2011)}]{Vijay2011}%
  \BibitemOpen
  \bibfield  {author} {\bibinfo {author} {\bibfnamefont {R.}~\bibnamefont
  {Vijay}}, \bibinfo {author} {\bibfnamefont {D.~H.}\ \bibnamefont
  {Slichter}},\ and\ \bibinfo {author} {\bibfnamefont {I.}~\bibnamefont
  {Siddiqi}},\ }\bibfield  {title} {\enquote {\bibinfo {title} {Observation of
  {Quantum} {Jumps} in a {Superconducting} {Artificial} {Atom}},}\ }\href
  {https://doi.org/10.1103/PhysRevLett.106.110502} {\bibfield  {journal}
  {\bibinfo  {journal} {Phys. Rev. Lett.}\ }\textbf {\bibinfo {volume} {106}},\
  \bibinfo {pages} {110502} (\bibinfo {year} {2011})}\BibitemShut {NoStop}%
\bibitem [{\citenamefont {Flurin}\ \emph {et~al.}(2012)\citenamefont {Flurin},
  \citenamefont {Roch}, \citenamefont {Mallet}, \citenamefont {Devoret},\ and\
  \citenamefont {Huard}}]{Flurin2012}%
  \BibitemOpen
  \bibfield  {author} {\bibinfo {author} {\bibfnamefont {E.}~\bibnamefont
  {Flurin}}, \bibinfo {author} {\bibfnamefont {N.}~\bibnamefont {Roch}},
  \bibinfo {author} {\bibfnamefont {F.}~\bibnamefont {Mallet}}, \bibinfo
  {author} {\bibfnamefont {M.~H.}\ \bibnamefont {Devoret}},\ and\ \bibinfo
  {author} {\bibfnamefont {B.}~\bibnamefont {Huard}},\ }\bibfield  {title}
  {\enquote {\bibinfo {title} {Generating {Entangled} {Microwave} {Radiation}
  {Over} {Two} {Transmission} {Lines}},}\ }\href
  {https://doi.org/10.1103/PhysRevLett.109.183901} {\bibfield  {journal}
  {\bibinfo  {journal} {Phys. Rev. Lett.}\ }\textbf {\bibinfo {volume} {109}},\
  \bibinfo {pages} {183901} (\bibinfo {year} {2012})}\BibitemShut {NoStop}%
\bibitem [{\citenamefont {Lin}\ \emph {et~al.}(2013)\citenamefont {Lin},
  \citenamefont {Inomata}, \citenamefont {Oliver}, \citenamefont {Koshino},
  \citenamefont {Nakamura}, \citenamefont {Tsai},\ and\ \citenamefont
  {Yamamoto}}]{Lin2013}%
  \BibitemOpen
  \bibfield  {author} {\bibinfo {author} {\bibfnamefont {Z.~R.}\ \bibnamefont
  {Lin}}, \bibinfo {author} {\bibfnamefont {K.}~\bibnamefont {Inomata}},
  \bibinfo {author} {\bibfnamefont {W.~D.}\ \bibnamefont {Oliver}}, \bibinfo
  {author} {\bibfnamefont {K.}~\bibnamefont {Koshino}}, \bibinfo {author}
  {\bibfnamefont {Y.}~\bibnamefont {Nakamura}}, \bibinfo {author}
  {\bibfnamefont {J.~S.}\ \bibnamefont {Tsai}},\ and\ \bibinfo {author}
  {\bibfnamefont {T.}~\bibnamefont {Yamamoto}},\ }\bibfield  {title} {\enquote
  {\bibinfo {title} {Single-shot readout of a superconducting flux qubit with a
  flux-driven {Josephson} parametric amplifier},}\ }\href
  {https://doi.org/10.1063/1.4821822} {\bibfield  {journal} {\bibinfo
  {journal} {Appl. Phys. Lett.}\ }\textbf {\bibinfo {volume} {103}},\ \bibinfo
  {pages} {132602} (\bibinfo {year} {2013})}\BibitemShut {NoStop}%
\bibitem [{\citenamefont {Vool}\ \emph {et~al.}(2016)\citenamefont {Vool},
  \citenamefont {Shankar}, \citenamefont {Mundhada}, \citenamefont {Ofek},
  \citenamefont {Narla}, \citenamefont {Sliwa}, \citenamefont {Zalys-Geller},
  \citenamefont {Liu}, \citenamefont {Frunzio}, \citenamefont {Schoelkopf},
  \citenamefont {Girvin},\ and\ \citenamefont {Devoret}}]{Vool2016}%
  \BibitemOpen
  \bibfield  {author} {\bibinfo {author} {\bibfnamefont {U.}~\bibnamefont
  {Vool}}, \bibinfo {author} {\bibfnamefont {S.}~\bibnamefont {Shankar}},
  \bibinfo {author} {\bibfnamefont {S.}~\bibnamefont {Mundhada}}, \bibinfo
  {author} {\bibfnamefont {N.}~\bibnamefont {Ofek}}, \bibinfo {author}
  {\bibfnamefont {A.}~\bibnamefont {Narla}}, \bibinfo {author} {\bibfnamefont
  {K.}~\bibnamefont {Sliwa}}, \bibinfo {author} {\bibfnamefont
  {E.}~\bibnamefont {Zalys-Geller}}, \bibinfo {author} {\bibfnamefont
  {Y.}~\bibnamefont {Liu}}, \bibinfo {author} {\bibfnamefont {L.}~\bibnamefont
  {Frunzio}}, \bibinfo {author} {\bibfnamefont {R.}~\bibnamefont {Schoelkopf}},
  \bibinfo {author} {\bibfnamefont {S.}~\bibnamefont {Girvin}},\ and\ \bibinfo
  {author} {\bibfnamefont {M.}~\bibnamefont {Devoret}},\ }\bibfield  {title}
  {\enquote {\bibinfo {title} {Continuous {Quantum} {Nondemolition}
  {Measurement} of the {Transverse} {Component} of a {Qubit}},}\ }\href
  {https://doi.org/10.1103/PhysRevLett.117.133601} {\bibfield  {journal}
  {\bibinfo  {journal} {Phys. Rev. Lett.}\ }\textbf {\bibinfo {volume} {117}},\
  \bibinfo {pages} {133601} (\bibinfo {year} {2016})}\BibitemShut {NoStop}%
\bibitem [{\citenamefont {Devoret}\ and\ \citenamefont
  {Schoelkopf}(2013)}]{Devoret2013}%
  \BibitemOpen
  \bibfield  {author} {\bibinfo {author} {\bibfnamefont {M.~H.}\ \bibnamefont
  {Devoret}}\ and\ \bibinfo {author} {\bibfnamefont {R.~J.}\ \bibnamefont
  {Schoelkopf}},\ }\bibfield  {title} {\enquote {\bibinfo {title}
  {Superconducting {Circuits} for {Quantum} {Information}: {An} {Outlook}},}\
  }\href {https://doi.org/10.1126/science.1231930} {\bibfield  {journal}
  {\bibinfo  {journal} {Science}\ }\textbf {\bibinfo {volume} {339}},\ \bibinfo
  {pages} {1169--1174} (\bibinfo {year} {2013})}\BibitemShut {NoStop}%
\bibitem [{\citenamefont {Yaakobi}\ \emph {et~al.}(2013)\citenamefont
  {Yaakobi}, \citenamefont {Friedland}, \citenamefont {Macklin},\ and\
  \citenamefont {Siddiqi}}]{Yaakobi2013}%
  \BibitemOpen
  \bibfield  {author} {\bibinfo {author} {\bibfnamefont {O.}~\bibnamefont
  {Yaakobi}}, \bibinfo {author} {\bibfnamefont {L.}~\bibnamefont {Friedland}},
  \bibinfo {author} {\bibfnamefont {C.}~\bibnamefont {Macklin}},\ and\ \bibinfo
  {author} {\bibfnamefont {I.}~\bibnamefont {Siddiqi}},\ }\bibfield  {title}
  {\enquote {\bibinfo {title} {Parametric amplification in {Josephson} junction
  embedded transmission lines},}\ }\href
  {https://doi.org/10.1103/PhysRevB.87.144301} {\bibfield  {journal} {\bibinfo
  {journal} {Phys. Rev. B}\ }\textbf {\bibinfo {volume} {87}},\ \bibinfo
  {pages} {144301} (\bibinfo {year} {2013})}\BibitemShut {NoStop}%
\bibitem [{\citenamefont {O’Brien}\ \emph {et~al.}(2014)\citenamefont
  {O’Brien}, \citenamefont {Macklin}, \citenamefont {Siddiqi},\ and\
  \citenamefont {Zhang}}]{OBrien2014}%
  \BibitemOpen
  \bibfield  {author} {\bibinfo {author} {\bibfnamefont {K.}~\bibnamefont
  {O’Brien}}, \bibinfo {author} {\bibfnamefont {C.}~\bibnamefont {Macklin}},
  \bibinfo {author} {\bibfnamefont {I.}~\bibnamefont {Siddiqi}},\ and\ \bibinfo
  {author} {\bibfnamefont {X.}~\bibnamefont {Zhang}},\ }\bibfield  {title}
  {\enquote {\bibinfo {title} {Resonant {Phase} {Matching} of {Josephson}
  {Junction} {Traveling} {Wave} {Parametric} {Amplifiers}},}\ }\href
  {https://doi.org/10.1103/PhysRevLett.113.157001} {\bibfield  {journal}
  {\bibinfo  {journal} {Phys. Rev. Lett.}\ }\textbf {\bibinfo {volume} {113}},\
  \bibinfo {pages} {157001} (\bibinfo {year} {2014})}\BibitemShut {NoStop}%
\bibitem [{\citenamefont {White}\ \emph {et~al.}(2015)\citenamefont {White},
  \citenamefont {Mutus}, \citenamefont {Hoi}, \citenamefont {Barends},
  \citenamefont {Campbell}, \citenamefont {Chen}, \citenamefont {Chen},
  \citenamefont {Chiaro}, \citenamefont {Dunsworth}, \citenamefont {Jeffrey},
  \citenamefont {Kelly}, \citenamefont {Megrant}, \citenamefont {Neill},
  \citenamefont {O'Malley}, \citenamefont {Roushan}, \citenamefont {Sank},
  \citenamefont {Vainsencher}, \citenamefont {Wenner}, \citenamefont
  {Chaudhuri}, \citenamefont {Gao},\ and\ \citenamefont
  {Martinis}}]{White2015}%
  \BibitemOpen
  \bibfield  {author} {\bibinfo {author} {\bibfnamefont {T.~C.}\ \bibnamefont
  {White}}, \bibinfo {author} {\bibfnamefont {J.~Y.}\ \bibnamefont {Mutus}},
  \bibinfo {author} {\bibfnamefont {I.-C.}\ \bibnamefont {Hoi}}, \bibinfo
  {author} {\bibfnamefont {R.}~\bibnamefont {Barends}}, \bibinfo {author}
  {\bibfnamefont {B.}~\bibnamefont {Campbell}}, \bibinfo {author}
  {\bibfnamefont {Y.}~\bibnamefont {Chen}}, \bibinfo {author} {\bibfnamefont
  {Z.}~\bibnamefont {Chen}}, \bibinfo {author} {\bibfnamefont {B.}~\bibnamefont
  {Chiaro}}, \bibinfo {author} {\bibfnamefont {A.}~\bibnamefont {Dunsworth}},
  \bibinfo {author} {\bibfnamefont {E.}~\bibnamefont {Jeffrey}}, \bibinfo
  {author} {\bibfnamefont {J.}~\bibnamefont {Kelly}}, \bibinfo {author}
  {\bibfnamefont {A.}~\bibnamefont {Megrant}}, \bibinfo {author} {\bibfnamefont
  {C.}~\bibnamefont {Neill}}, \bibinfo {author} {\bibfnamefont {P.~J.~J.}\
  \bibnamefont {O'Malley}}, \bibinfo {author} {\bibfnamefont {P.}~\bibnamefont
  {Roushan}}, \bibinfo {author} {\bibfnamefont {D.}~\bibnamefont {Sank}},
  \bibinfo {author} {\bibfnamefont {A.}~\bibnamefont {Vainsencher}}, \bibinfo
  {author} {\bibfnamefont {J.}~\bibnamefont {Wenner}}, \bibinfo {author}
  {\bibfnamefont {S.}~\bibnamefont {Chaudhuri}}, \bibinfo {author}
  {\bibfnamefont {J.}~\bibnamefont {Gao}},\ and\ \bibinfo {author}
  {\bibfnamefont {J.~M.}\ \bibnamefont {Martinis}},\ }\bibfield  {title}
  {\enquote {\bibinfo {title} {Traveling wave parametric amplifier with
  {Josephson} junctions using minimal resonator phase matching},}\ }\href
  {https://doi.org/10.1063/1.4922348} {\bibfield  {journal} {\bibinfo
  {journal} {Appl. Phys. Lett.}\ }\textbf {\bibinfo {volume} {106}},\ \bibinfo
  {pages} {242601} (\bibinfo {year} {2015})}\BibitemShut {NoStop}%
\bibitem [{\citenamefont {Macklin}\ \emph {et~al.}(2015)\citenamefont
  {Macklin}, \citenamefont {O’Brien}, \citenamefont {Hover}, \citenamefont
  {Schwartz}, \citenamefont {Bolkhovsky}, \citenamefont {Zhang}, \citenamefont
  {Oliver},\ and\ \citenamefont {Siddiqi}}]{Macklin2015}%
  \BibitemOpen
  \bibfield  {author} {\bibinfo {author} {\bibfnamefont {C.}~\bibnamefont
  {Macklin}}, \bibinfo {author} {\bibfnamefont {K.}~\bibnamefont {O’Brien}},
  \bibinfo {author} {\bibfnamefont {D.}~\bibnamefont {Hover}}, \bibinfo
  {author} {\bibfnamefont {M.~E.}\ \bibnamefont {Schwartz}}, \bibinfo {author}
  {\bibfnamefont {V.}~\bibnamefont {Bolkhovsky}}, \bibinfo {author}
  {\bibfnamefont {X.}~\bibnamefont {Zhang}}, \bibinfo {author} {\bibfnamefont
  {W.~D.}\ \bibnamefont {Oliver}},\ and\ \bibinfo {author} {\bibfnamefont
  {I.}~\bibnamefont {Siddiqi}},\ }\bibfield  {title} {\enquote {\bibinfo
  {title} {A near–quantum-limited {Josephson} traveling-wave parametric
  amplifier},}\ }\href {https://doi.org/10.1126/science.aaa8525} {\bibfield
  {journal} {\bibinfo  {journal} {Science}\ }\textbf {\bibinfo {volume}
  {350}},\ \bibinfo {pages} {307--310} (\bibinfo {year} {2015})}\BibitemShut
  {NoStop}%
\bibitem [{\citenamefont {Bell}\ and\ \citenamefont
  {Samolov}(2015)}]{Bell-Samolov2015}%
  \BibitemOpen
  \bibfield  {author} {\bibinfo {author} {\bibfnamefont {M.}~\bibnamefont
  {Bell}}\ and\ \bibinfo {author} {\bibfnamefont {A.}~\bibnamefont {Samolov}},\
  }\bibfield  {title} {\enquote {\bibinfo {title} {Traveling-{Wave}
  {Parametric} {Amplifier} {Based} on a {Chain} of {Coupled} {Asymmetric}
  {SQUIDs}},}\ }\href {https://doi.org/10.1103/PhysRevApplied.4.024014}
  {\bibfield  {journal} {\bibinfo  {journal} {Phys. Rev. Applied}\ }\textbf
  {\bibinfo {volume} {4}},\ \bibinfo {pages} {024014} (\bibinfo {year}
  {2015})}\BibitemShut {NoStop}%
\bibitem [{\citenamefont {Zorin}(2016)}]{Zorin2016}%
  \BibitemOpen
  \bibfield  {author} {\bibinfo {author} {\bibfnamefont {A.}~\bibnamefont
  {Zorin}},\ }\bibfield  {title} {\enquote {\bibinfo {title} {Josephson
  {Traveling}-{Wave} {Parametric} {Amplifier} with {Three}-{Wave} {Mixing}},}\
  }\href {https://doi.org/10.1103/PhysRevApplied.6.034006} {\bibfield
  {journal} {\bibinfo  {journal} {Phys. Rev. Applied}\ }\textbf {\bibinfo
  {volume} {6}},\ \bibinfo {pages} {034006} (\bibinfo {year}
  {2016})}\BibitemShut {NoStop}%
\bibitem [{\citenamefont {Zorin}\ \emph {et~al.}(2017)\citenamefont {Zorin},
  \citenamefont {Khabipov}, \citenamefont {Dietel},\ and\ \citenamefont
  {Dolata}}]{Zorin2017}%
  \BibitemOpen
  \bibfield  {author} {\bibinfo {author} {\bibfnamefont {A.~B.}\ \bibnamefont
  {Zorin}}, \bibinfo {author} {\bibfnamefont {M.}~\bibnamefont {Khabipov}},
  \bibinfo {author} {\bibfnamefont {J.}~\bibnamefont {Dietel}},\ and\ \bibinfo
  {author} {\bibfnamefont {R.}~\bibnamefont {Dolata}},\ }\bibfield  {title}
  {\enquote {\bibinfo {title} {Traveling-{Wave} {Parametric} {Amplifier}
  {Based} on {Three}-{Wave} {Mixing} in a {Josephson} {Metamaterial}},}\ }in\
  \href {https://doi.org/10.1109/ISEC.2017.8314196} {\emph {\bibinfo
  {booktitle} {2017 16th {International} {Superconductive} {Electronics}
  {Conference} ({ISEC})}}}\ (\bibinfo {year} {2017})\ pp.\ \bibinfo {pages}
  {1--3}\BibitemShut {NoStop}%
\bibitem [{\citenamefont {Zhang}\ \emph {et~al.}(2017)\citenamefont {Zhang},
  \citenamefont {Huang}, \citenamefont {Gershenson},\ and\ \citenamefont
  {Bell}}]{WenyuanZhang2017}%
  \BibitemOpen
  \bibfield  {author} {\bibinfo {author} {\bibfnamefont {W.}~\bibnamefont
  {Zhang}}, \bibinfo {author} {\bibfnamefont {W.}~\bibnamefont {Huang}},
  \bibinfo {author} {\bibfnamefont {M.}~\bibnamefont {Gershenson}},\ and\
  \bibinfo {author} {\bibfnamefont {M.}~\bibnamefont {Bell}},\ }\bibfield
  {title} {\enquote {\bibinfo {title} {Josephson {Metamaterial} with a {Widely}
  {Tunable} {Positive} or {Negative} {Kerr} {Constant}},}\ }\href
  {https://doi.org/10.1103/PhysRevApplied.8.051001} {\bibfield  {journal}
  {\bibinfo  {journal} {Phys. Rev. Applied}\ }\textbf {\bibinfo {volume} {8}},\
  \bibinfo {pages} {051001} (\bibinfo {year} {2017})}\BibitemShut {NoStop}%
\bibitem [{\citenamefont {Miano}\ and\ \citenamefont
  {Mukhanov}(2019)}]{Miano2018}%
  \BibitemOpen
  \bibfield  {author} {\bibinfo {author} {\bibfnamefont {A.}~\bibnamefont
  {Miano}}\ and\ \bibinfo {author} {\bibfnamefont {O.~A.}\ \bibnamefont
  {Mukhanov}},\ }\bibfield  {title} {\enquote {\bibinfo {title} {Symmetric
  {Traveling} {Wave} {Parametric} {Amplifier}},}\ }\href
  {https://doi.org/10.1109/TASC.2019.2904699} {\bibfield  {journal} {\bibinfo
  {journal} {IEEE Trans. Appl. Supercond.}\ }\textbf {\bibinfo {volume} {29}},\
  \bibinfo {pages} {1--6} (\bibinfo {year} {2019})}\BibitemShut {NoStop}%
\bibitem [{\citenamefont {Zorin}(2019)}]{Zorin2019}%
  \BibitemOpen
  \bibfield  {author} {\bibinfo {author} {\bibfnamefont {A.}~\bibnamefont
  {Zorin}},\ }\bibfield  {title} {\enquote {\bibinfo {title} {Flux-{Driven}
  {Josephson} {Traveling}-{Wave} {Parametric} {Amplifier}},}\ }\href
  {https://doi.org/10.1103/PhysRevApplied.12.044051} {\bibfield  {journal}
  {\bibinfo  {journal} {Phys. Rev. Applied}\ }\textbf {\bibinfo {volume}
  {12}},\ \bibinfo {pages} {044051} (\bibinfo {year} {2019})}\BibitemShut
  {NoStop}%
\bibitem [{\citenamefont {Planat}\ \emph {et~al.}(2020)\citenamefont {Planat},
  \citenamefont {Ranadive}, \citenamefont {Dassonneville}, \citenamefont
  {Puertas~Martínez}, \citenamefont {Léger}, \citenamefont {Naud},
  \citenamefont {Buisson}, \citenamefont {Hasch-Guichard}, \citenamefont
  {Basko},\ and\ \citenamefont {Roch}}]{Planat2020}%
  \BibitemOpen
  \bibfield  {author} {\bibinfo {author} {\bibfnamefont {L.}~\bibnamefont
  {Planat}}, \bibinfo {author} {\bibfnamefont {A.}~\bibnamefont {Ranadive}},
  \bibinfo {author} {\bibfnamefont {R.}~\bibnamefont {Dassonneville}}, \bibinfo
  {author} {\bibfnamefont {J.}~\bibnamefont {Puertas~Martínez}}, \bibinfo
  {author} {\bibfnamefont {S.}~\bibnamefont {Léger}}, \bibinfo {author}
  {\bibfnamefont {C.}~\bibnamefont {Naud}}, \bibinfo {author} {\bibfnamefont
  {O.}~\bibnamefont {Buisson}}, \bibinfo {author} {\bibfnamefont
  {W.}~\bibnamefont {Hasch-Guichard}}, \bibinfo {author} {\bibfnamefont
  {D.~M.}\ \bibnamefont {Basko}},\ and\ \bibinfo {author} {\bibfnamefont
  {N.}~\bibnamefont {Roch}},\ }\bibfield  {title} {\enquote {\bibinfo {title}
  {Photonic-{Crystal} {Josephson} {Traveling}-{Wave} {Parametric}
  {Amplifier}},}\ }\href {https://doi.org/10.1103/PhysRevX.10.021021}
  {\bibfield  {journal} {\bibinfo  {journal} {Phys. Rev. X}\ }\textbf {\bibinfo
  {volume} {10}},\ \bibinfo {pages} {021021} (\bibinfo {year}
  {2020})}\BibitemShut {NoStop}%
\bibitem [{\citenamefont {Ranadive}\ \emph {et~al.}(2022)\citenamefont
  {Ranadive}, \citenamefont {Esposito}, \citenamefont {Planat}, \citenamefont
  {Bonet}, \citenamefont {Naud}, \citenamefont {Buisson}, \citenamefont
  {Guichard},\ and\ \citenamefont {Roch}}]{Ranadive2022}%
  \BibitemOpen
  \bibfield  {author} {\bibinfo {author} {\bibfnamefont {A.}~\bibnamefont
  {Ranadive}}, \bibinfo {author} {\bibfnamefont {M.}~\bibnamefont {Esposito}},
  \bibinfo {author} {\bibfnamefont {L.}~\bibnamefont {Planat}}, \bibinfo
  {author} {\bibfnamefont {E.}~\bibnamefont {Bonet}}, \bibinfo {author}
  {\bibfnamefont {C.}~\bibnamefont {Naud}}, \bibinfo {author} {\bibfnamefont
  {O.}~\bibnamefont {Buisson}}, \bibinfo {author} {\bibfnamefont
  {W.}~\bibnamefont {Guichard}},\ and\ \bibinfo {author} {\bibfnamefont
  {N.}~\bibnamefont {Roch}},\ }\bibfield  {title} {\enquote {\bibinfo {title}
  {Kerr reversal in {Josephson} meta-material and traveling wave parametric
  amplification},}\ }\href {https://doi.org/10.1038/s41467-022-29375-5}
  {\bibfield  {journal} {\bibinfo  {journal} {Nature Communications}\ }\textbf
  {\bibinfo {volume} {13}},\ \bibinfo {pages} {1737} (\bibinfo {year}
  {2022})}\BibitemShut {NoStop}%
\bibitem [{\citenamefont {Zorin}(2021)}]{Zorin2021}%
  \BibitemOpen
  \bibfield  {author} {\bibinfo {author} {\bibfnamefont {A.~B.}\ \bibnamefont
  {Zorin}},\ }\bibfield  {title} {\enquote {\bibinfo {title}
  {Quasi-phasematching in a poled {Josephson} traveling-wave parametric
  amplifier with three-wave mixing},}\ }\href
  {https://doi.org/10.1063/5.0050787} {\bibfield  {journal} {\bibinfo
  {journal} {Appl. Phys. Lett.}\ }\textbf {\bibinfo {volume} {118}},\ \bibinfo
  {pages} {222601} (\bibinfo {year} {2021})}\BibitemShut {NoStop}%
\bibitem [{\citenamefont {Josephson}(1962)}]{Josephson1962}%
  \BibitemOpen
  \bibfield  {author} {\bibinfo {author} {\bibfnamefont {B.~D.}\ \bibnamefont
  {Josephson}},\ }\bibfield  {title} {\enquote {\bibinfo {title} {Possible new
  effects in superconductive tunnelling},}\ }\href
  {https://doi.org/10.1016/0031-9163(62)91369-0} {\bibfield  {journal}
  {\bibinfo  {journal} {Phys. Lett.}\ }\textbf {\bibinfo {volume} {1}},\
  \bibinfo {pages} {251--253} (\bibinfo {year} {1962})}\BibitemShut {NoStop}%
\bibitem [{\citenamefont {Agrawal}(2001)}]{Agrawal}%
  \BibitemOpen
  \bibfield  {author} {\bibinfo {author} {\bibfnamefont {G.~P.}\ \bibnamefont
  {Agrawal}},\ }\href@noop {} {\emph {\bibinfo {title} {Applications of
  nonlinear fiber optics}}},\ Optics and photonics\ (\bibinfo  {publisher}
  {Academic Press},\ \bibinfo {address} {San Diego},\ \bibinfo {year}
  {2001})\BibitemShut {NoStop}%
\bibitem [{\citenamefont {Tien}(1958)}]{Tien1958}%
  \BibitemOpen
  \bibfield  {author} {\bibinfo {author} {\bibfnamefont {P.~K.}\ \bibnamefont
  {Tien}},\ }\bibfield  {title} {\enquote {\bibinfo {title} {Parametric
  {Amplification} and {Frequency} {Mixing} in {Propagating} {Circuits}},}\
  }\href {https://doi.org/10.1063/1.1723440} {\bibfield  {journal} {\bibinfo
  {journal} {J. Appl. Phys.}\ }\textbf {\bibinfo {volume} {29}},\ \bibinfo
  {pages} {1347--1357} (\bibinfo {year} {1958})}\BibitemShut {NoStop}%
\bibitem [{\citenamefont {Cullen}(1960)}]{Cullen1960}%
  \BibitemOpen
  \bibfield  {author} {\bibinfo {author} {\bibfnamefont {A.}~\bibnamefont
  {Cullen}},\ }\bibfield  {title} {\enquote {\bibinfo {title} {Theory of the
  travelling-wave parametric amplifier},}\ }\href
  {https://doi.org/10.1049/pi-b-2.1960.0085} {\bibfield  {journal} {\bibinfo
  {journal} {Proc. IEE Part B: Electron. and Communication Eng.}\ }\textbf
  {\bibinfo {volume} {107}},\ \bibinfo {pages} {101} (\bibinfo {year}
  {1960})}\BibitemShut {NoStop}%
\bibitem [{\citenamefont {Sivak}\ \emph {et~al.}(2019)\citenamefont {Sivak},
  \citenamefont {Frattini}, \citenamefont {Joshi}, \citenamefont
  {Lingenfelter}, \citenamefont {Shankar},\ and\ \citenamefont
  {Devoret}}]{Sivak2019}%
  \BibitemOpen
  \bibfield  {author} {\bibinfo {author} {\bibfnamefont {V.}~\bibnamefont
  {Sivak}}, \bibinfo {author} {\bibfnamefont {N.}~\bibnamefont {Frattini}},
  \bibinfo {author} {\bibfnamefont {V.}~\bibnamefont {Joshi}}, \bibinfo
  {author} {\bibfnamefont {A.}~\bibnamefont {Lingenfelter}}, \bibinfo {author}
  {\bibfnamefont {S.}~\bibnamefont {Shankar}},\ and\ \bibinfo {author}
  {\bibfnamefont {M.}~\bibnamefont {Devoret}},\ }\bibfield  {title} {\enquote
  {\bibinfo {title} {Kerr-{Free} {Three}-{Wave} {Mixing} in {Superconducting}
  {Quantum} {Circuits}},}\ }\href
  {https://doi.org/10.1103/PhysRevApplied.11.054060} {\bibfield  {journal}
  {\bibinfo  {journal} {Phys. Rev. Applied}\ }\textbf {\bibinfo {volume}
  {11}},\ \bibinfo {pages} {054060} (\bibinfo {year} {2019})}\BibitemShut
  {NoStop}%
\bibitem [{\citenamefont {Frattini}\ \emph {et~al.}(2017)\citenamefont
  {Frattini}, \citenamefont {Vool}, \citenamefont {Shankar}, \citenamefont
  {Narla}, \citenamefont {Sliwa},\ and\ \citenamefont
  {Devoret}}]{Frattini2017}%
  \BibitemOpen
  \bibfield  {author} {\bibinfo {author} {\bibfnamefont {N.~E.}\ \bibnamefont
  {Frattini}}, \bibinfo {author} {\bibfnamefont {U.}~\bibnamefont {Vool}},
  \bibinfo {author} {\bibfnamefont {S.}~\bibnamefont {Shankar}}, \bibinfo
  {author} {\bibfnamefont {A.}~\bibnamefont {Narla}}, \bibinfo {author}
  {\bibfnamefont {K.~M.}\ \bibnamefont {Sliwa}},\ and\ \bibinfo {author}
  {\bibfnamefont {M.~H.}\ \bibnamefont {Devoret}},\ }\bibfield  {title}
  {\enquote {\bibinfo {title} {3-wave mixing {Josephson} dipole element},}\
  }\href {https://doi.org/10.1063/1.4984142} {\bibfield  {journal} {\bibinfo
  {journal} {Appl. Phys. Lett.}\ }\textbf {\bibinfo {volume} {110}},\ \bibinfo
  {pages} {222603} (\bibinfo {year} {2017})}\BibitemShut {NoStop}%
\bibitem [{\citenamefont {Frattini}\ \emph {et~al.}(2018)\citenamefont
  {Frattini}, \citenamefont {Sivak}, \citenamefont {Lingenfelter},
  \citenamefont {Shankar},\ and\ \citenamefont {Devoret}}]{Frattini2018}%
  \BibitemOpen
  \bibfield  {author} {\bibinfo {author} {\bibfnamefont {N.~E.}\ \bibnamefont
  {Frattini}}, \bibinfo {author} {\bibfnamefont {V.~V.}\ \bibnamefont {Sivak}},
  \bibinfo {author} {\bibfnamefont {A.}~\bibnamefont {Lingenfelter}}, \bibinfo
  {author} {\bibfnamefont {S.}~\bibnamefont {Shankar}},\ and\ \bibinfo {author}
  {\bibfnamefont {M.~H.}\ \bibnamefont {Devoret}},\ }\bibfield  {title}
  {\enquote {\bibinfo {title} {Optimizing the {Nonlinearity} and {Dissipation}
  of a {SNAIL} {Parametric} {Amplifier} for {Dynamic} {Range}},}\ }\href
  {https://doi.org/10.1103/PhysRevApplied.10.054020} {\bibfield  {journal}
  {\bibinfo  {journal} {Phys. Rev. Applied}\ }\textbf {\bibinfo {volume}
  {10}},\ \bibinfo {pages} {054020} (\bibinfo {year} {2018})}\BibitemShut
  {NoStop}%
\bibitem [{\citenamefont {Dixon}\ \emph {et~al.}(2020)\citenamefont {Dixon},
  \citenamefont {Dunstan}, \citenamefont {Long}, \citenamefont {Williams},
  \citenamefont {Meeson},\ and\ \citenamefont {Shelly}}]{Dixon2020}%
  \BibitemOpen
  \bibfield  {author} {\bibinfo {author} {\bibfnamefont {T.}~\bibnamefont
  {Dixon}}, \bibinfo {author} {\bibfnamefont {J.}~\bibnamefont {Dunstan}},
  \bibinfo {author} {\bibfnamefont {G.}~\bibnamefont {Long}}, \bibinfo {author}
  {\bibfnamefont {J.}~\bibnamefont {Williams}}, \bibinfo {author}
  {\bibfnamefont {P.}~\bibnamefont {Meeson}},\ and\ \bibinfo {author}
  {\bibfnamefont {C.}~\bibnamefont {Shelly}},\ }\bibfield  {title} {\enquote
  {\bibinfo {title} {Capturing {Complex} {Behavior} in {Josephson}
  {Traveling}-{Wave} {Parametric} {Amplifiers}},}\ }\href
  {https://doi.org/10.1103/PhysRevApplied.14.034058} {\bibfield  {journal}
  {\bibinfo  {journal} {Phys. Rev. Applied}\ }\textbf {\bibinfo {volume}
  {14}},\ \bibinfo {pages} {034058} (\bibinfo {year} {2020})}\BibitemShut
  {NoStop}%
\bibitem [{\citenamefont {Malnou}\ \emph {et~al.}(2021)\citenamefont {Malnou},
  \citenamefont {Vissers}, \citenamefont {Wheeler}, \citenamefont {Aumentado},
  \citenamefont {Hubmayr}, \citenamefont {Ullom},\ and\ \citenamefont
  {Gao}}]{Malnou2021}%
  \BibitemOpen
  \bibfield  {author} {\bibinfo {author} {\bibfnamefont {M.}~\bibnamefont
  {Malnou}}, \bibinfo {author} {\bibfnamefont {M.}~\bibnamefont {Vissers}},
  \bibinfo {author} {\bibfnamefont {J.}~\bibnamefont {Wheeler}}, \bibinfo
  {author} {\bibfnamefont {J.}~\bibnamefont {Aumentado}}, \bibinfo {author}
  {\bibfnamefont {J.}~\bibnamefont {Hubmayr}}, \bibinfo {author} {\bibfnamefont
  {J.}~\bibnamefont {Ullom}},\ and\ \bibinfo {author} {\bibfnamefont
  {J.}~\bibnamefont {Gao}},\ }\bibfield  {title} {\enquote {\bibinfo {title}
  {Three-{Wave} {Mixing} {Kinetic} {Inductance} {Traveling}-{Wave} {Amplifier}
  with {Near}-{Quantum}-{Limited} {Noise} {Performance}},}\ }\href
  {https://doi.org/10.1103/PRXQuantum.2.010302} {\bibfield  {journal} {\bibinfo
   {journal} {PRX Quantum}\ }\textbf {\bibinfo {volume} {2}},\ \bibinfo {pages}
  {010302} (\bibinfo {year} {2021})}\BibitemShut {NoStop}%
\bibitem [{\citenamefont {Perelshtein}\ \emph {et~al.}(2021)\citenamefont
  {Perelshtein}, \citenamefont {Petrovnin}, \citenamefont {Vesterinen},
  \citenamefont {Raja}, \citenamefont {Lilja}, \citenamefont {Will},
  \citenamefont {Savin}, \citenamefont {Simbierowicz}, \citenamefont
  {Jabdaraghi}, \citenamefont {Lehtinen}, \citenamefont {Grönberg},
  \citenamefont {Hassel}, \citenamefont {Prunnila}, \citenamefont {Govenius},
  \citenamefont {Paraoanu},\ and\ \citenamefont {Hakonen}}]{Perelshtein21}%
  \BibitemOpen
  \bibfield  {author} {\bibinfo {author} {\bibfnamefont {M.}~\bibnamefont
  {Perelshtein}}, \bibinfo {author} {\bibfnamefont {K.}~\bibnamefont
  {Petrovnin}}, \bibinfo {author} {\bibfnamefont {V.}~\bibnamefont
  {Vesterinen}}, \bibinfo {author} {\bibfnamefont {S.~H.}\ \bibnamefont
  {Raja}}, \bibinfo {author} {\bibfnamefont {I.}~\bibnamefont {Lilja}},
  \bibinfo {author} {\bibfnamefont {M.}~\bibnamefont {Will}}, \bibinfo {author}
  {\bibfnamefont {A.}~\bibnamefont {Savin}}, \bibinfo {author} {\bibfnamefont
  {S.}~\bibnamefont {Simbierowicz}}, \bibinfo {author} {\bibfnamefont
  {R.}~\bibnamefont {Jabdaraghi}}, \bibinfo {author} {\bibfnamefont
  {J.}~\bibnamefont {Lehtinen}}, \bibinfo {author} {\bibfnamefont
  {L.}~\bibnamefont {Grönberg}}, \bibinfo {author} {\bibfnamefont
  {J.}~\bibnamefont {Hassel}}, \bibinfo {author} {\bibfnamefont
  {M.}~\bibnamefont {Prunnila}}, \bibinfo {author} {\bibfnamefont
  {J.}~\bibnamefont {Govenius}}, \bibinfo {author} {\bibfnamefont
  {S.}~\bibnamefont {Paraoanu}},\ and\ \bibinfo {author} {\bibfnamefont
  {P.}~\bibnamefont {Hakonen}},\ }\href {http://arxiv.org/abs/2111.06145}
  {\enquote {\bibinfo {title} {Broadband continuous variable entanglement
  generation using {Kerr}-free {Josephson} metamaterial},}\ } (\bibinfo {year}
  {2021}),\ \Eprint {https://arxiv.org/abs/2111.06145} {arXiv:2111.06145
  [cond-mat.supr-con]} \BibitemShut {NoStop}%
\bibitem [{\citenamefont {Clarke}\ and\ \citenamefont
  {Braginski}(2004)}]{ClarkeBraginski2004}%
  \BibitemOpen
  \bibinfo {editor} {\bibfnamefont {J.}~\bibnamefont {Clarke}}\ and\ \bibinfo
  {editor} {\bibfnamefont {A.~I.}\ \bibnamefont {Braginski}},\ eds.,\ \href
  {https://doi.org/10.1002/3527603646} {\emph {\bibinfo {title} {The {SQUID}
  {Handbook}: {Fundamentals} and {Technology} of {SQUIDs} and {SQUID}
  {Systems}}}},\ \bibinfo {edition} {1st}\ ed.\ (\bibinfo  {publisher}
  {Wiley},\ \bibinfo {year} {2004})\BibitemShut {NoStop}%
\bibitem [{\citenamefont {Zhao}\ and\ \citenamefont
  {Withington}(2021)}]{Zhao2021}%
  \BibitemOpen
  \bibfield  {author} {\bibinfo {author} {\bibfnamefont {S.}~\bibnamefont
  {Zhao}}\ and\ \bibinfo {author} {\bibfnamefont {S.}~\bibnamefont
  {Withington}},\ }\bibfield  {title} {\enquote {\bibinfo {title} {Quantum
  analysis of second-order effects in superconducting travelling-wave
  parametric amplifiers},}\ }\href {https://doi.org/10.1088/1361-6463/ac0b74}
  {\bibfield  {journal} {\bibinfo  {journal} {Journal of Physics D: Applied
  Physics}\ }\textbf {\bibinfo {volume} {54}},\ \bibinfo {pages} {365303}
  (\bibinfo {year} {2021})}\BibitemShut {NoStop}%
\bibitem [{\citenamefont {Ho~Eom}\ \emph {et~al.}(2012)\citenamefont {Ho~Eom},
  \citenamefont {Day}, \citenamefont {LeDuc},\ and\ \citenamefont
  {Zmuidzinas}}]{Eom2012}%
  \BibitemOpen
  \bibfield  {author} {\bibinfo {author} {\bibfnamefont {B.}~\bibnamefont
  {Ho~Eom}}, \bibinfo {author} {\bibfnamefont {P.~K.}\ \bibnamefont {Day}},
  \bibinfo {author} {\bibfnamefont {H.~G.}\ \bibnamefont {LeDuc}},\ and\
  \bibinfo {author} {\bibfnamefont {J.}~\bibnamefont {Zmuidzinas}},\ }\bibfield
   {title} {\enquote {\bibinfo {title} {A wideband, low-noise superconducting
  amplifier with high dynamic range},}\ }\href
  {https://doi.org/10.1038/nphys2356} {\bibfield  {journal} {\bibinfo
  {journal} {Nat. Phys.}\ }\textbf {\bibinfo {volume} {8}},\ \bibinfo {pages}
  {623--627} (\bibinfo {year} {2012})}\BibitemShut {NoStop}%
\bibitem [{\citenamefont {Vissers}\ \emph {et~al.}(2016)\citenamefont
  {Vissers}, \citenamefont {Erickson}, \citenamefont {Ku}, \citenamefont
  {Vale}, \citenamefont {Wu}, \citenamefont {Hilton},\ and\ \citenamefont
  {Pappas}}]{Vissers2016}%
  \BibitemOpen
  \bibfield  {author} {\bibinfo {author} {\bibfnamefont {M.~R.}\ \bibnamefont
  {Vissers}}, \bibinfo {author} {\bibfnamefont {R.~P.}\ \bibnamefont
  {Erickson}}, \bibinfo {author} {\bibfnamefont {H.-S.}\ \bibnamefont {Ku}},
  \bibinfo {author} {\bibfnamefont {L.}~\bibnamefont {Vale}}, \bibinfo {author}
  {\bibfnamefont {X.}~\bibnamefont {Wu}}, \bibinfo {author} {\bibfnamefont
  {G.~C.}\ \bibnamefont {Hilton}},\ and\ \bibinfo {author} {\bibfnamefont
  {D.~P.}\ \bibnamefont {Pappas}},\ }\bibfield  {title} {\enquote {\bibinfo
  {title} {Low-noise kinetic inductance traveling-wave amplifier using
  three-wave mixing},}\ }\href {https://doi.org/10.1063/1.4937922} {\bibfield
  {journal} {\bibinfo  {journal} {Appl. Phys. Lett.}\ }\textbf {\bibinfo
  {volume} {108}},\ \bibinfo {pages} {012601} (\bibinfo {year}
  {2016})}\BibitemShut {NoStop}%
\bibitem [{\citenamefont {Erickson}\ and\ \citenamefont
  {Pappas}(2017)}]{Erickson2017}%
  \BibitemOpen
  \bibfield  {author} {\bibinfo {author} {\bibfnamefont {R.~P.}\ \bibnamefont
  {Erickson}}\ and\ \bibinfo {author} {\bibfnamefont {D.~P.}\ \bibnamefont
  {Pappas}},\ }\bibfield  {title} {\enquote {\bibinfo {title} {Theory of
  multiwave mixing within the superconducting kinetic-inductance traveling-wave
  amplifier},}\ }\href {https://doi.org/10.1103/PhysRevB.95.104506} {\bibfield
  {journal} {\bibinfo  {journal} {Phys. Rev. B}\ }\textbf {\bibinfo {volume}
  {95}},\ \bibinfo {pages} {104506} (\bibinfo {year} {2017})}\BibitemShut
  {NoStop}%
\bibitem [{\citenamefont {Roudsari}\ \emph {et~al.}(2022)\citenamefont
  {Roudsari}, \citenamefont {Shiri}, \citenamefont {Nilsson}, \citenamefont
  {Tancredi}, \citenamefont {Osman}, \citenamefont {Svensson}, \citenamefont
  {Kudra}, \citenamefont {Rommel}, \citenamefont {Bylander}, \citenamefont
  {Shumeiko},\ and\ \citenamefont {Delsing}}]{Roudsari_arxiv2022}%
  \BibitemOpen
  \bibfield  {author} {\bibinfo {author} {\bibfnamefont {A.~F.}\ \bibnamefont
  {Roudsari}}, \bibinfo {author} {\bibfnamefont {D.}~\bibnamefont {Shiri}},
  \bibinfo {author} {\bibfnamefont {H.~R.}\ \bibnamefont {Nilsson}}, \bibinfo
  {author} {\bibfnamefont {G.}~\bibnamefont {Tancredi}}, \bibinfo {author}
  {\bibfnamefont {A.}~\bibnamefont {Osman}}, \bibinfo {author} {\bibfnamefont
  {I.-M.}\ \bibnamefont {Svensson}}, \bibinfo {author} {\bibfnamefont
  {M.}~\bibnamefont {Kudra}}, \bibinfo {author} {\bibfnamefont
  {M.}~\bibnamefont {Rommel}}, \bibinfo {author} {\bibfnamefont
  {J.}~\bibnamefont {Bylander}}, \bibinfo {author} {\bibfnamefont
  {V.}~\bibnamefont {Shumeiko}},\ and\ \bibinfo {author} {\bibfnamefont
  {P.}~\bibnamefont {Delsing}},\ }\bibfield  {title} {\enquote {\bibinfo
  {title} {Three-wave mixing traveling-wave parametric amplifier with periodic
  variation of the circuit parameters},}\ }\href
  {https://doi.org/10.48550/ARXIV.2209.07551} {\  (\bibinfo {year} {2022}),\
  10.48550/ARXIV.2209.07551}\BibitemShut {NoStop}%
\bibitem [{\citenamefont {Nilsson}\ \emph {et~al.}(2022)\citenamefont
  {Nilsson}, \citenamefont {Roudsari}, \citenamefont {Shiri}, \citenamefont
  {Delsing},\ and\ \citenamefont {Shumeiko}}]{Nillson_arxiv2022}%
  \BibitemOpen
  \bibfield  {author} {\bibinfo {author} {\bibfnamefont {H.~R.}\ \bibnamefont
  {Nilsson}}, \bibinfo {author} {\bibfnamefont {A.~F.}\ \bibnamefont
  {Roudsari}}, \bibinfo {author} {\bibfnamefont {D.}~\bibnamefont {Shiri}},
  \bibinfo {author} {\bibfnamefont {P.}~\bibnamefont {Delsing}},\ and\ \bibinfo
  {author} {\bibfnamefont {V.}~\bibnamefont {Shumeiko}},\ }\bibfield  {title}
  {\enquote {\bibinfo {title} {A high gain travelling-wave parametric amplifier
  based on three-wave mixing},}\ }\href
  {https://doi.org/10.48550/ARXIV.2205.07758} {\  (\bibinfo {year} {2022}),\
  10.48550/ARXIV.2205.07758}\BibitemShut {NoStop}%
\bibitem [{\citenamefont {Kronig}\ and\ \citenamefont
  {Penney}(1931)}]{KronigPenney1931}%
  \BibitemOpen
  \bibfield  {author} {\bibinfo {author} {\bibfnamefont {R.~D.~L.}\
  \bibnamefont {Kronig}}\ and\ \bibinfo {author} {\bibfnamefont {W.~G.}\
  \bibnamefont {Penney}},\ }\bibfield  {title} {\enquote {\bibinfo {title}
  {Quantum mechanics of electrons in crystal lattices},}\ }\href
  {https://doi.org/10.1098/rspa.1931.0019} {\bibfield  {journal} {\bibinfo
  {journal} {Proc. R. Soc. London Series A}\ }\textbf {\bibinfo {volume}
  {130}},\ \bibinfo {pages} {499--513} (\bibinfo {year} {1931})}\BibitemShut
  {NoStop}%
\bibitem [{\citenamefont {Kittel}(2005)}]{Kittel2004}%
  \BibitemOpen
  \bibfield  {author} {\bibinfo {author} {\bibfnamefont {C.}~\bibnamefont
  {Kittel}},\ }\href@noop {} {\emph {\bibinfo {title} {Introduction to solid
  state physics}}},\ \bibinfo {edition} {8th}\ ed.\ (\bibinfo  {publisher}
  {Wiley},\ \bibinfo {address} {Hoboken, NJ},\ \bibinfo {year}
  {2005})\BibitemShut {NoStop}%
\bibitem [{\citenamefont {Mackay}\ and\ \citenamefont
  {Lakhtakia}(2020)}]{Mackay2020}%
  \BibitemOpen
  \bibfield  {author} {\bibinfo {author} {\bibfnamefont {T.~G.}\ \bibnamefont
  {Mackay}}\ and\ \bibinfo {author} {\bibfnamefont {A.}~\bibnamefont
  {Lakhtakia}},\ }\href {https://doi.org/10.2200/S00993ED1V01Y202002EMA001}
  {\emph {\bibinfo {title} {The {Transfer}-{Matrix} {Method} in
  {Electromagnetics} and {Optics}}}},\ Synthesis {Lectures} on
  {Electromagnetics}\ (\bibinfo  {publisher} {Morgan \& Claypool},\ \bibinfo
  {address} {San Rafael},\ \bibinfo {year} {2020})\BibitemShut {NoStop}%
\bibitem [{WRs()}]{WRspice}%
  \BibitemOpen
  \href {http://www.wrcad.com/wrspice.html} {\enquote {\bibinfo {title}
  {{Whiteley} {Research} {Inc.}, {WRspice} {Circuit} {Simulator}},}\ }\bibinfo
  {howpublished} {http://www.wrcad.com/wrspice.html}\BibitemShut {NoStop}%
\bibitem [{\citenamefont {{Ó~Peatáin}}\ \emph {et~al.}(2021)\citenamefont
  {{Ó~Peatáin}}, \citenamefont {Dixon}, \citenamefont {Meeson}, \citenamefont
  {Williams}, \citenamefont {Kafanov},\ and\ \citenamefont
  {Pashkin}}]{OPeatain2021preprint}%
  \BibitemOpen
  \bibfield  {author} {\bibinfo {author} {\bibfnamefont {S.}~\bibnamefont
  {{Ó~Peatáin}}}, \bibinfo {author} {\bibfnamefont {T.}~\bibnamefont
  {Dixon}}, \bibinfo {author} {\bibfnamefont {P.~J.}\ \bibnamefont {Meeson}},
  \bibinfo {author} {\bibfnamefont {J.}~\bibnamefont {Williams}}, \bibinfo
  {author} {\bibfnamefont {S.}~\bibnamefont {Kafanov}},\ and\ \bibinfo {author}
  {\bibfnamefont {Y.~A.}\ \bibnamefont {Pashkin}},\ }\href@noop {} {\enquote
  {\bibinfo {title} {The {Effect} of {Parameter} {Variations} on the
  {Performance} of the {Josephson} {Travelling} {Wave} {Parametric}
  {Amplifiers}},}\ } (\bibinfo {year} {2021}),\ \Eprint
  {https://arxiv.org/abs/2112.07766} {arXiv:2112.07766 [cond-mat.supr-con]}
  \BibitemShut {NoStop}%
\bibitem [{\citenamefont {Whiteley}(1991)}]{whiteley1991}%
  \BibitemOpen
  \bibfield  {author} {\bibinfo {author} {\bibfnamefont {S.}~\bibnamefont
  {Whiteley}},\ }\bibfield  {title} {\enquote {\bibinfo {title} {Josephson
  junctions in {SPICE3}},}\ }\href {https://doi.org/10.1109/20.133816}
  {\bibfield  {journal} {\bibinfo  {journal} {IEEE Trans. Magn.}\ }\textbf
  {\bibinfo {volume} {27}},\ \bibinfo {pages} {2902--2905} (\bibinfo {year}
  {1991})}\BibitemShut {NoStop}%
\bibitem [{\citenamefont {Jewett}(1982)}]{jewett1982josephson}%
  \BibitemOpen
  \bibfield  {author} {\bibinfo {author} {\bibfnamefont {R.~E.}\ \bibnamefont
  {Jewett}},\ }\bibfield  {title} {\enquote {\bibinfo {title} {Josephson
  junctions in {SPICE} {2G5}},}\ }\href@noop {} {\bibfield  {journal} {\bibinfo
   {journal} {Electronics Research Laboratory internal memoranda, University of
  {California}, {Berkeley}, {CA}}\ } (\bibinfo {year} {1982})}\BibitemShut
  {NoStop}%
\bibitem [{\citenamefont {Pozar}(2012)}]{Pozar2012}%
  \BibitemOpen
  \bibfield  {author} {\bibinfo {author} {\bibfnamefont {D.~M.}\ \bibnamefont
  {Pozar}},\ }\href@noop {} {\emph {\bibinfo {title} {Microwave
  engineering}}},\ \bibinfo {edition} {4th}\ ed.\ (\bibinfo  {publisher}
  {Wiley},\ \bibinfo {address} {Hoboken, NJ},\ \bibinfo {year}
  {2012})\BibitemShut {NoStop}%
\bibitem [{\citenamefont {Qiu}\ \emph {et~al.}(2022)\citenamefont {Qiu},
  \citenamefont {Grimsmo}, \citenamefont {Peng}, \citenamefont {Kannan},
  \citenamefont {Lienhard}, \citenamefont {Sung}, \citenamefont {Krantz},
  \citenamefont {Bolkhovsky}, \citenamefont {Calusine}, \citenamefont {Kim},
  \citenamefont {Melville}, \citenamefont {Niedzielski}, \citenamefont {Yoder},
  \citenamefont {Schwartz}, \citenamefont {Orlando}, \citenamefont {Siddiqi},
  \citenamefont {Gustavsson}, \citenamefont {O'Brien},\ and\ \citenamefont
  {Oliver}}]{Qiu2022preprint}%
  \BibitemOpen
  \bibfield  {author} {\bibinfo {author} {\bibfnamefont {J.~Y.}\ \bibnamefont
  {Qiu}}, \bibinfo {author} {\bibfnamefont {A.}~\bibnamefont {Grimsmo}},
  \bibinfo {author} {\bibfnamefont {K.}~\bibnamefont {Peng}}, \bibinfo {author}
  {\bibfnamefont {B.}~\bibnamefont {Kannan}}, \bibinfo {author} {\bibfnamefont
  {B.}~\bibnamefont {Lienhard}}, \bibinfo {author} {\bibfnamefont
  {Y.}~\bibnamefont {Sung}}, \bibinfo {author} {\bibfnamefont {P.}~\bibnamefont
  {Krantz}}, \bibinfo {author} {\bibfnamefont {V.}~\bibnamefont {Bolkhovsky}},
  \bibinfo {author} {\bibfnamefont {G.}~\bibnamefont {Calusine}}, \bibinfo
  {author} {\bibfnamefont {D.}~\bibnamefont {Kim}}, \bibinfo {author}
  {\bibfnamefont {A.}~\bibnamefont {Melville}}, \bibinfo {author}
  {\bibfnamefont {B.~M.}\ \bibnamefont {Niedzielski}}, \bibinfo {author}
  {\bibfnamefont {J.}~\bibnamefont {Yoder}}, \bibinfo {author} {\bibfnamefont
  {M.~E.}\ \bibnamefont {Schwartz}}, \bibinfo {author} {\bibfnamefont {T.~P.}\
  \bibnamefont {Orlando}}, \bibinfo {author} {\bibfnamefont {I.}~\bibnamefont
  {Siddiqi}}, \bibinfo {author} {\bibfnamefont {S.}~\bibnamefont {Gustavsson}},
  \bibinfo {author} {\bibfnamefont {K.~P.}\ \bibnamefont {O'Brien}},\ and\
  \bibinfo {author} {\bibfnamefont {W.~D.}\ \bibnamefont {Oliver}},\
  }\href@noop {} {\enquote {\bibinfo {title} {Broadband {Squeezed} {Microwaves}
  and {Amplification} with a {Josephson} {Traveling}-{Wave} {Parametric}
  {Amplifier}},}\ } (\bibinfo {year} {2022}),\ \Eprint
  {https://arxiv.org/abs/2201.11261} {arXiv:2201.11261 [quant-ph]} \BibitemShut
  {NoStop}%
\bibitem [{\citenamefont {Southwell}(1989)}]{Southwell1989}%
  \BibitemOpen
  \bibfield  {author} {\bibinfo {author} {\bibfnamefont {W.~H.}\ \bibnamefont
  {Southwell}},\ }\bibfield  {title} {\enquote {\bibinfo {title} {Using
  apodization functions to reduce sidelobes in rugate filters},}\ }\href
  {https://doi.org/10.1364/AO.28.005091} {\bibfield  {journal} {\bibinfo
  {journal} {Applied Optics}\ }\textbf {\bibinfo {volume} {28}},\ \bibinfo
  {pages} {5091--5094} (\bibinfo {year} {1989})}\BibitemShut {NoStop}%
\bibitem [{\citenamefont {Abu-Safia}, \citenamefont {Al-Sharif},\ and\
  \citenamefont {Abu~Aljarayesh}(1993)}]{AbuSafia1993}%
  \BibitemOpen
  \bibfield  {author} {\bibinfo {author} {\bibfnamefont {H.~A.}\ \bibnamefont
  {Abu-Safia}}, \bibinfo {author} {\bibfnamefont {A.~I.}\ \bibnamefont
  {Al-Sharif}},\ and\ \bibinfo {author} {\bibfnamefont {I.~O.}\ \bibnamefont
  {Abu~Aljarayesh}},\ }\bibfield  {title} {\enquote {\bibinfo {title} {Rugate
  filter sidelobe suppression using half-apodization},}\ }\href
  {https://doi.org/10.1364/AO.32.004831} {\bibfield  {journal} {\bibinfo
  {journal} {Applied Optics}\ }\textbf {\bibinfo {volume} {32}},\ \bibinfo
  {pages} {4831} (\bibinfo {year} {1993})}\BibitemShut {NoStop}%
\bibitem [{\citenamefont {Peng}\ \emph {et~al.}(2022)\citenamefont {Peng},
  \citenamefont {Naghiloo}, \citenamefont {Wang}, \citenamefont {Cunningham},
  \citenamefont {Ye},\ and\ \citenamefont {O’Brien}}]{Peng2022}%
  \BibitemOpen
  \bibfield  {author} {\bibinfo {author} {\bibfnamefont {K.}~\bibnamefont
  {Peng}}, \bibinfo {author} {\bibfnamefont {M.}~\bibnamefont {Naghiloo}},
  \bibinfo {author} {\bibfnamefont {J.}~\bibnamefont {Wang}}, \bibinfo {author}
  {\bibfnamefont {G.~D.}\ \bibnamefont {Cunningham}}, \bibinfo {author}
  {\bibfnamefont {Y.}~\bibnamefont {Ye}},\ and\ \bibinfo {author}
  {\bibfnamefont {K.~P.}\ \bibnamefont {O’Brien}},\ }\bibfield  {title}
  {\enquote {\bibinfo {title} {Floquet-{Mode} {Traveling}-{Wave} {Parametric}
  {Amplifiers}},}\ }\href {https://doi.org/10.1103/PRXQuantum.3.020306}
  {\bibfield  {journal} {\bibinfo  {journal} {PRX Quantum}\ }\textbf {\bibinfo
  {volume} {3}},\ \bibinfo {pages} {020306} (\bibinfo {year}
  {2022})}\BibitemShut {NoStop}%
\bibitem [{\citenamefont {Greco}\ \emph {et~al.}(2021)\citenamefont {Greco},
  \citenamefont {Fasolo}, \citenamefont {Meda}, \citenamefont {Callegaro},\
  and\ \citenamefont {Enrico}}]{Greco2021}%
  \BibitemOpen
  \bibfield  {author} {\bibinfo {author} {\bibfnamefont {A.}~\bibnamefont
  {Greco}}, \bibinfo {author} {\bibfnamefont {L.}~\bibnamefont {Fasolo}},
  \bibinfo {author} {\bibfnamefont {A.}~\bibnamefont {Meda}}, \bibinfo {author}
  {\bibfnamefont {L.}~\bibnamefont {Callegaro}},\ and\ \bibinfo {author}
  {\bibfnamefont {E.}~\bibnamefont {Enrico}},\ }\bibfield  {title} {\enquote
  {\bibinfo {title} {Quantum model for rf-squid-based metamaterials enabling
  three-wave mixing and four-wave mixing traveling-wave parametric
  amplification},}\ }\href {https://doi.org/10.1103/PhysRevB.104.184517}
  {\bibfield  {journal} {\bibinfo  {journal} {Phys. Rev. B}\ }\textbf {\bibinfo
  {volume} {104}},\ \bibinfo {pages} {184517} (\bibinfo {year}
  {2021})}\BibitemShut {NoStop}%
\bibitem [{\citenamefont {Fasolo}\ \emph {et~al.}(2021)\citenamefont {Fasolo},
  \citenamefont {Greco}, \citenamefont {Enrico}, \citenamefont {Illuminati},
  \citenamefont {Lo~Franco}, \citenamefont {Vitali},\ and\ \citenamefont
  {Livreri}}]{Fasolo2021}%
  \BibitemOpen
  \bibfield  {author} {\bibinfo {author} {\bibfnamefont {L.}~\bibnamefont
  {Fasolo}}, \bibinfo {author} {\bibfnamefont {A.}~\bibnamefont {Greco}},
  \bibinfo {author} {\bibfnamefont {E.}~\bibnamefont {Enrico}}, \bibinfo
  {author} {\bibfnamefont {F.}~\bibnamefont {Illuminati}}, \bibinfo {author}
  {\bibfnamefont {R.}~\bibnamefont {Lo~Franco}}, \bibinfo {author}
  {\bibfnamefont {D.}~\bibnamefont {Vitali}},\ and\ \bibinfo {author}
  {\bibfnamefont {P.}~\bibnamefont {Livreri}},\ }\bibfield  {title} {\enquote
  {\bibinfo {title} {Josephson {Traveling} {Wave} {Parametric} {Amplifiers} as
  non-classical light source for {Microwave} {Quantum} {Illumination}},}\
  }\href {https://doi.org/10.1016/j.measen.2021.100349} {\bibfield  {journal}
  {\bibinfo  {journal} {Measurement: Sensors}\ }\textbf {\bibinfo {volume}
  {18}},\ \bibinfo {pages} {100349} (\bibinfo {year} {2021})}\BibitemShut
  {NoStop}%
\bibitem [{\citenamefont {Gaydamachenko}\ and\ \citenamefont
  {Kissling}(2022)}]{Zenodo-rep}%
  \BibitemOpen
  \bibfield  {author} {\bibinfo {author} {\bibfnamefont {V.}~\bibnamefont
  {Gaydamachenko}}\ and\ \bibinfo {author} {\bibfnamefont {C.}~\bibnamefont
  {Kissling}},\ }\bibfield  {title} {\enquote {\bibinfo {title} {Numerical
  analysis of a three-wave-mixing {Josephson} traveling-wave parametric
  amplifier with engineered dispersion loadings - data and {WRspice}
  netlist},}\ }\href {https://doi.org/https://doi.org/10.5281/zenodo.7092947}
  {\  (\bibinfo {year} {2022}),\
  https://doi.org/10.5281/zenodo.7092947}\BibitemShut {NoStop}%
\end{thebibliography}%

\end{document}